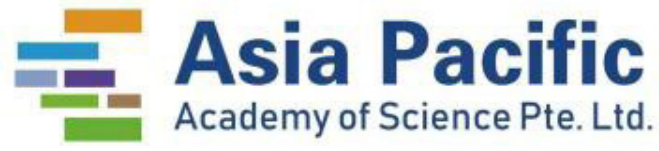

Review

# Metaverse in Smart Cities: Transforming Urban Life and Governance

**Ioanna Chatzopoulou[1], Paraskevi Tsoutsa[2], Panos Fitsilis[1*]**

[1] Department of Business Administration University of Thessaly, Greece
[2] Department of Accounting and Finance University of Thessaly, Greece
*** Corresponding author:** Panos Fitsilis, pfitsilis@gmail.com

**Abstract:** The integration of metaverse technologies within Smart Cities (SCs) is transforming urban governance and citizen engagement. Despite the increasing academic and industry interest, research on the practical applications of the metaverse in SCs remains fragmented. This study addresses this gap through a systematic literature review on how metaverse-driven solutions impact economic transformation, governance, mobility, sustainability, and social interactions in urban environments. Motivated by the growing demand for immersive, data-driven, and participatory SC solutions, this research applies Giffinger's SC model to evaluate metaverse integration across key SC dimensions. The study synthesizes findings from existing applications and case studies, such as Metaverse Seoul, Dubai's Metaverse Strategy, Virtual Helsinki, and Tampere's CitiVerse initiative, to illustrate the diverse ways in which cities are leveraging metaverse technologies. These applications demonstrate the metaverse's potential in digital governance, Artificial Intelligence (AI)-driven urban planning, e-participation, transportation optimization, and climate resilience strategies. This research contributes to the field by providing a comprehensive framework for understanding the benefits and challenges of metaverse-driven SC models. The findings suggest that while metaverse adoption in SCs presents significant advantages in efficiency, participation, and innovation, it also entails challenges related to technological accessibility, governance frameworks, and security measures that must be addressed for broad uptake. The study's impact extends to policymakers, urban planners, and technology developers by offering strategic insights for responsible and inclusive metaverse adoption. Ultimately, this study provides a structured roadmap for integrating metaverse technologies into smart urban ecosystems, ensuring their long-term viability, accessibility, and effectiveness in shaping the cities of the future.

## 1. Introduction

The SC paradigm embodies a multidimensional framework that strategically incorporates advanced technologies into urban governance and infrastructure to optimize service delivery, sustainability, and the quality of urban life [1,2]. The COVID-19 pandemic has significantly accelerated this transformation, facilitating the integration of metaverse technologies within the SC ecosystem. This shift has led to the emergence of a digitally interconnected urban model, the metaverse-SC, that leverages AI, Extended Reality (XR), the Internet of Things (IoT), Big Data, and other cutting-edge technologies to enhance urban experiences and redefine governance [3,4]. By merging the physical and virtual realms, these technologies enable advanced urban planning, data-driven decision-making, and enhanced civic participation [4,5]. More than just a technological evolution, metaverse integration is reshaping cultural norms, governance models, and economic structures within urban ecosystems [6]. While the potential of metaverse-enhanced SCs has been widely recognized, their practical applications and long-term implications remain underexplored in scholarly discourse [3,7].

Existing literature has examined the metaverse's influence on societal dynamics [8,9], technological advances [10–12], education [13–15], cybersecurity [16–18], healthcare [19], tourism [20], and consumer behavior [21–23]. However, research investigating the direct application of metaverse technologies in real-world SCs remains limited, especially empirical studies that explore how these technologies are being integrated into urban infrastructure, sustainability initiatives, and participatory governance. This study addresses this gap by undertaking a systematic literature review of metaverse adoption in SCs, with an emphasis on real-world applications and their implications for sustainable, inclusive, and resilient urban environments. Guided by the foundational framework established by Giffinger et al. [24], this research combines a systematic literature review with case study analysis to elucidate how the metaverse refigures essential components of SCs. By examining metaverse-driven transformations across smart economy, smart people, smart governance, smart mobility, smart environment, and smart living, the study identifies emerging trends, practical applications, and underlying technological enablers shaping urban digitalization.

The primary objectives of this research are as follows:

(i) To investigate the defining characteristics and real-world implementations of metaverse technologies in SCs through case studies, highlighting their transformative impact on urban governance, infrastructure, and citizen engagement.

(ii) To assess the implications of metaverse integration for addressing urban challenges, optimizing service delivery, and enhancing civic participation, particularly through Digital Twins (DTs), AI-driven simulations, and immersive virtual collaboration tools.

(iii) To analyze the CitiVerse concept as a digital extension of SCs, evaluating its capacity to create interconnected urban DTs that drive innovation, inclusivity, and enhanced social, economic, and governance functionalities.

(iv) To identify key challenges and critical risks and to recommend actionable countermeasures toward responsible metaverse adoption.

By accomplishing these objectives, this study contributes to the ongoing discourse on metaverse-enabled SCs by offering insights into best practices, potential risks, and policy considerations. Specifically, (i) the study's conceptual/theoretical contribution lies in integrating the CitiVerse model into the SC architecture, aligning core metaverse capabilities with Giffinger's 6 SC pillars; (ii) methodologically, it employs a Preferred Reporting Items for Systematic Reviews and Meta-Analyses (PRISMA)-based review combined with cross-city cases to explain variation across governance and data-rights regimes; (iii) empirically, it offers a comparative view of city metaverse deployments across economy, governance, mobility, culture, and environment, and provides a strategic roadmap with actionable guidance for policymakers and urban designers; (iv) in terms of novelty, it moves beyond education-and consumer-focused work, bridging the SC adoption gap and providing a reusable reference model with decision-oriented guidance for inclusive, responsible metaverse integration.

The remainder of the article is structured as follows: Section 2 reviews smart urbanism through the lens of Giffinger's SC model [24], while exploring its intersection with the metaverse. Section 3 delineates the research methodology and data collection processes employed in the review. Section 4 reports the results, aligning CitiVerse with the 6 pillars of SCs and illustrating applications with cross-city cases. Section 5 discusses key risks and opportunities and distills practical and research implications, highlighting the study's contributions and directions for responsible metaverse-SC integration. Finally, Section 6 concludes the study.

## 2. Background

### 2.1. SC key dimensions and areas of focus

The European Commission's Statistical Office (Eurostat) and UN-Habitat define a city as an urban center with at least 50,000 residents, assessed through spatial, statistical, and census data. In contrast, the term "Smart City" lacks a universal definition, leading to diverse interpretations across disciplines [25,26]. Historically, the SC concept emerged in the mid-1800s to describe efficiently managed, self-governing cities in the American West. By the 1990s, it evolved to encompass smart growth and sustainable urbanization, addressing challenges such as overcrowding, pollution, climate change, and crime [26,27]. With the urban population share projected to exceed 50 percent of the global total by 2050, SC discussions have gained urgency and complexity [28].

A defining feature of SCs is the integration of Information and Communication Technology (ICT), shaping adaptable, multi-layered urban systems [27]. Since 2005, major corporations, such as IBM, ABB, HP, Siemens, Ericsson, and Cisco, have developed advanced systems for smart urban operations, deploying IoT, cloud computing, and sensor networks [15,28]. However, the techno-centric model has faced criticism, prompting a shift toward citizen-centric, sustainable, and socially innovative approaches [29].

This evolution has moved SCs from first-generation, top-down, techno-economic models to more theoretical and ontological frameworks, and further to hybrid approaches that emphasize bottom-up governance, civic participation, and social inclusion [30,31]. Contemporary SCs are now conceived as collaborative ecosystems in which human and non-human actors engage in dialogue and negotiation to shape future urban development [25,32]. Despite varied interpretations, common SC characteristics include technology integration, standardization, transparency, efficiency, e-service delivery, innovation, sustainability, and public-private partnerships [33]. These elements define the multifaceted nature of modern SC urbanism, as **Table 1** illustrates.

**Table 1.** Comparative approaches to defining the SC concept.

| Key SC dimensions | Core focus & contributions | Researcher |
|---|---|---|
| **Technological, human, and institutional dimensions** | Introduces a three-dimensional framework that highlights the interplay between technology, human factors, and institutional structures in SCs. Focuses on urban innovation as a function of governance and policymaking. | Nam & Pardo [34] |
| **Viability, workability, and sustainability** | Advocates a playable-city approach, where urban planning incorporates creative and participatory dimensions. Highlights sustainability as a key factor in evaluating SC success. | Leorke [25] |
| **Triple-helix model (university, industry, government)** | Introduces a neo-evolutionary model that emphasizes knowledge-based economic growth through university-industry-government partnerships. Forms the foundation for later multi-stakeholder SC models. | Leydesdorff & Deakin [35] |

Continuation Table:

| Key SC dimensions | Core focus & contributions | Researcher |
|---|---|---|
| **Quadruple-helix model (government, industry, academia, civil society)** | Expands the triple-helix model by incorporating civil society as a crucial stakeholder in SC development. Highlights the role of collaboration and synergy among 4 sectors to drive urban transformation. | Kuzior & Kuzior [36] |
| **Penta-helix model (public, private, academia, civic society, and social entrepreneurs/activists)** | Proposes a multi-stakeholder governance model that extends previous helix models by integrating activists and social entrepreneurs as key agents in SC innovation. Supports the democratization of SC initiatives. | Calzada [37] |
| **Six-dimensional SC model (ICT, entrepreneurship, inclusive strategies, social capital, economic and environmental considerations)** | Presents an interdisciplinary framework integrating ICT-driven urban development, economic factors, and social inclusivity. One of the first SC models to incorporate both economic and environmental dimensions. | Caragliu et al. [38] |
| **E-participation and digital democracy** | Highlights citizen e-participation as a cornerstone of SC governance. Advocates for digital platforms that enhance citizen engagement in policymaking and urban development. | Benton & Cropf [39] |
| **Six-dimensional SC model (smart economy, smart mobility, smart environment, smart people, smart living, smart governance)** | Establishes a quantifiable assessment framework with 31 factors and 74 indicators to measure SC performance. Remains one of the most referenced models in SC research. | Giffinger et al. [24] |
| **SC indicators model (45 indicators in 6 categories based on Giffinger's model)** | Expands Giffinger's model by introducing additional indicators for more precise SC evaluation. Focuses on data-driven urban management and performance tracking. | Napitupulu et al. [40] |
| **SC ecosystem (governance, service innovation, partnership formation, proactiveness, infrastructure integration)** | Defines SCs as dynamic ecosystems that require urban openness, technological flexibility, and strong governance structures to remain adaptive. Stresses collaborative partnerships between various urban stakeholders. | Ooms et al. [41] |

The comparative analysis of the aforementioned SC models reveals a significant evolution from technology-driven frameworks to more human-centered, participatory approaches. Early models prioritized economic efficiency, infrastructure, and digital integration [24,38], whereas later frameworks emphasize citizen-participatory governance, inclusivity, and sustainability [37,39]. This shift is evident in the transition from the triple-helix (university-industry-government) model [35] to the quadruple-helix [36] and penta-helix [37], which expands multi-stakeholder engagement to include civil society and social entrepreneurs. However, the integration of digital governance and AI-enabled technologies remains uneven; some models [40,41] stress data-driven decision-making, while others [25] foreground urban viability with limited attention to digital transformation. Although Giffinger's six-dimensional SC model remains widely used, newer frameworks refine SC assessment indicators [40]. E-participation [39] and multi-stakeholder collaboration [37] are increasingly integral to citizen-led urban transformation.

A holistic, adaptive SC model should combine penta-helix collaboration, AI-driven governance, expanded digital inclusion, robust citizen e-participation, and climate-resilient strategies. Ultimately, the future of SCs hinges on fusing data-driven technologies with inclusive, citizen-oriented governance to deliver long-term urban resilience, innovation, and socio-environmental sustainability. Unlike concepts such as Digital City, Intelligent City, or Ubiquitous City, SCs emphasize human-centric urbanism and civic participation [42,43].

However, persistent challenges and critical risks—infrastructure gaps, limited digital literacy, cybersecurity and privacy concerns, public skepticism, and the high costs of immersive XR (VR/AR)—must be addressed to ensure that technological advances translate into inclusive, resilient, and sustainable urban environments [44–47].

## 2.2. Giffinger's SC model

The SC model by Giffinger et al. [24] is a foundational framework for evaluating urban sustainability, competitiveness, and governance. It defines 6 core dimensions, namely smart economy, smart people, smart governance, smart mobility, smart environment, and smart living, structured into 31 factors and 74 indicators, as shown in **Figure 1**. This data-driven methodology provides policymakers and researchers with a transparent, standardized tool for assessing SC effectiveness, first applied in Vienna and later adapted in Italy [48] and South Korea [49], confirming its applicability across diverse urban contexts. The model articulates a proactive urban strategy that integrates stakeholder perspectives, sustainability goals, social inclusion, and effective governance, making it an international reference for SC development [50].

Giffinger's model delineates essential SC qualities, including flexibility, adaptability, synergy, individuality, self-decisiveness, and strategic governance, emphasizing cross-sector collaboration, ICT integration, educational systems, and skilled inhabitants. By leveraging advanced digital infrastructure, SCs enhance mobility, sustainability, energy efficiency, and safety, reinforcing evidence-informed urban resilience and governance. Ultimately, Giffinger's model remains a cornerstone in SC research, bridging technology, data analytics, and participatory governance to shape future-ready, sustainable urban environments.

Each dimension of the model includes specific indicators:

**Smart Economy** consists of 6 factors and 12 indicators. Innovation is assessed through Research and Development (R&D) expenditure as a share of Gross Domestic Product (GDP), patents, and workforce specialization. Entrepreneurship is reflected in self-employment rates and business start-ups. Labor market flexibility is gauged through GDP per employee, employment statistics, and job transition rates. International integration examines multinational headquarters, trade volume, and air traffic. Economic branding evaluates the city's attractiveness for investors and businesses. A 7th factor, transformative capability, remains important but is less validated in literature.

**Smart People** analyzes human capital through 7 factors and 15 indicators. Education and lifelong learning are measured through book-borrowing rates (library circulation), university presence, and participation in lifelong learning programs. Social and ethnic diversity captures migrant integration and cultural inclusion. Openness and participation reflect political engagement, volunteerism, and democratic participation, while creativity and flexibility are indicated by employment in cultural and creative industries.

**Smart Governance** includes 4 factors and 9 indicators. Participatory decision-making assesses citizen engagement in political processes, while public and social services evaluate accessibility and effectiveness. Transparent governance considers accountability measures, while strategic political vision assesses long-term policy frameworks.

**Smart Mobility** emphasizes logistics and infrastructure, comprising 4

factors and 9 indicators. Sustainable transport examines public-transit use, eco-friendly mobility options, and traffic safety. Local accessibility measures intra-city connectivity, while national and international accessibility examines air, rail, and road networks [51]. Additionally, ICT infrastructure scrutinizes the availability of devices and internet connectivity.

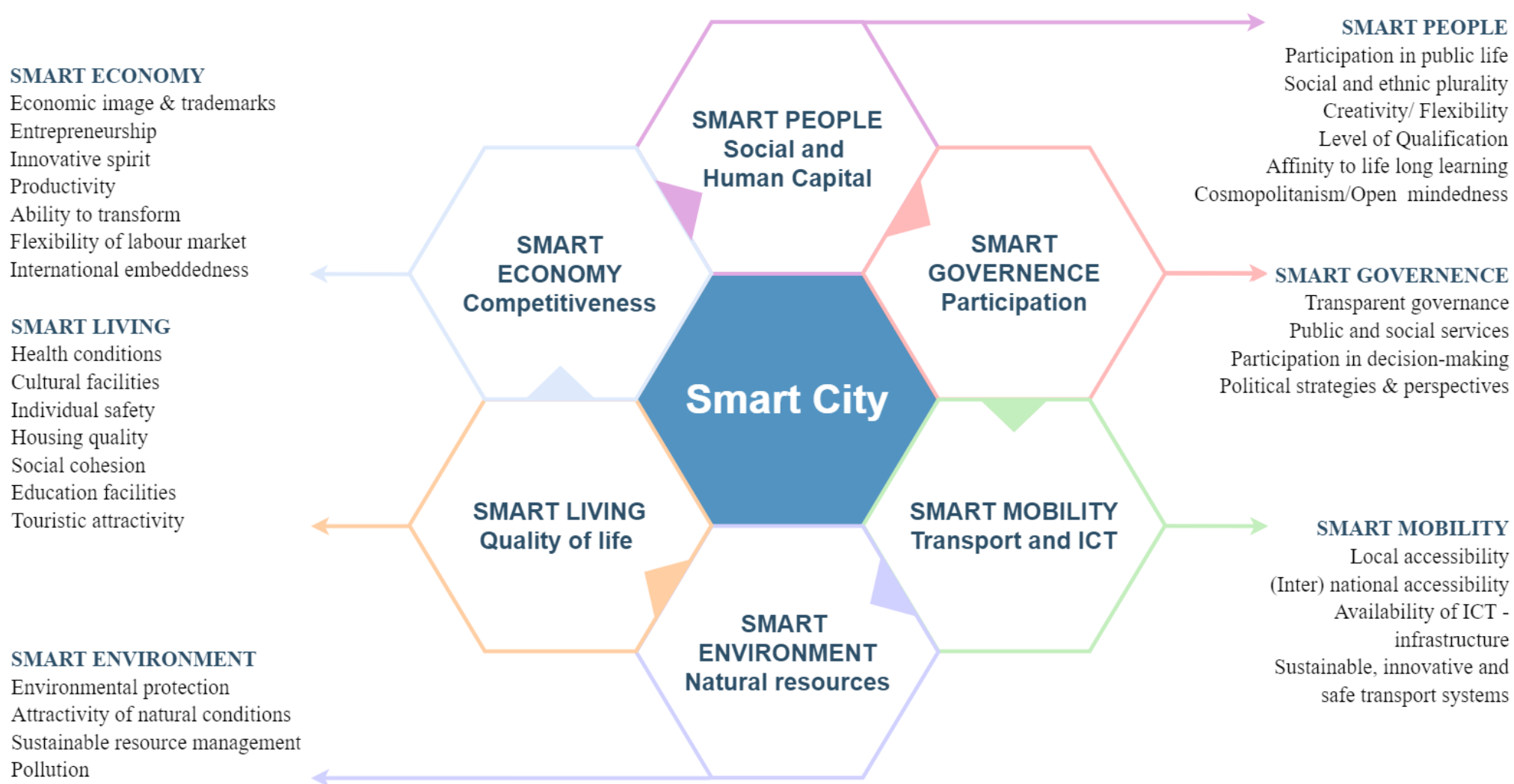

**Figure 1.** Giffinger's six-dimensional model of SCs includes smart economy, smart people, smart governance, smart mobility, smart environment, and smart living (adapted by the author).

**Smart Environment** consists of 4 factors and 9 indicators. Resource management considers water and energy efficiency. Pollution control assesses air quality and public health impacts. Natural assets cover green spaces and biodiversity, while climate adaptation strategies reflect resilience initiatives.

**Smart Living** encompasses 7 factors and 20 indicators. Personal safety is assessed through crime rates and public perceptions of security. Housing quality reflects affordability and living conditions. Cultural facilities consider attendance at cultural venues, while educational facilities examine access to education and student attainment. Healthcare is measured using hospital capacity and life expectancy. Tourist attractions reflect the city's appeal to visitors (e.g., annual overnight stays). Social cohesion is measured through poverty levels and community solidarity.

Giffinger's model is selected for this research due to its structured and replicable methodology, offering a comprehensive framework for assessing urban functionality, sustainability, and governance. Its proven adaptability across diverse socio-economic and cultural contexts—including Vienna, Italy, China, and South Korea [49,50,52]—solidifies its continued relevance in contemporary SC discourse. Beyond methodological rigor, the model integrates stakeholder engagement, sustainability priorities, and strategic governance, which are critical for effective SC implementation. By embedding these components, it enhances urban resilience, digital transformation, and citizen-centered innovation, establishing a robust benchmarking tool for policymakers and urban planners. Furthermore, its quantifiable indicators enable precise comparative analysis, performance tracking, and targeted policy refinement, thereby supporting evidence-based and optimized decision-making for sustainable urban development.

Serving as the theoretical and empirical backbone of this research, the model reinforces best practices in SC planning and guides cities toward technologically advanced, socially inclusive, and environmentally resilient futures.

### 2.3. Metaverse overview

Originating in Neal Stephenson's "Snow Crash" [53], the metaverse envisions a fully immersive 3D digital universe where users interact as avatars in real time [3,11,16]. This concept is not new; it has been echoed by science-fiction authors, including Isaac Asimov and Arthur C. Clarke, and has permeated popular culture through works like "Ready Player One" [54,55]. While initially speculative, it has since evolved into a disruptive force in education, commerce, governance, healthcare, and entertainment [16]. The term gained mainstream attention in 2021 with Facebook's rebranding to Meta, which positioned the metaverse as a central element of societal reconstruction in the wake of the COVID-19 pandemic [56].

The International Telecommunication Union's (ITU) technical report on the metaverse synthesizes 150 existing definitions and defines the metaverse as "an integrative ecosystem of virtual worlds offering immersive experiences to users and modifying preexisting value and creating new value from economic, environmental, social, and cultural perspectives" [57,58].

Unlike conventional digital environments, the metaverse is a decentralized and persistent space where digital assets, smart contracts, and blockchain technologies facilitate transactions, collaboration, and social interaction [59,60]. The term itself, derived from the Greek prefix "meta" (beyond) and "verse" (universe), signifies its ambition to create a digital realm unconstrained by physical limitations [61]. As a result, it enables users to establish virtual identities, participate in social and economic systems, and engage in novel forms of communication, governance, and commerce [44,62].

Hudson-Smith [63] (p.343) identifies 7 prerequisites for the metaverse, including robust connectivity for seamless data exchange, advanced hardware (headsets, haptic devices) for immersive interaction, decentralized ecosystems (blockchain) for transparency, high-fidelity 3D visualization for realism, digital-asset ownership for empowerment, diverse user experiences, and virtual equivalents of real-world activities. Key factors, such as low latency, scalability, interoperability, security, privacy, and energy efficiency, are essential for integrating the metaverse into everyday life [9]. Additionally, its persistent and programmable operation allows developers to schedule events and users to interact without external constraints [29,64].

The metaverse is built on cutting-edge technologies [15,16,29,59,65–70] including:

- AI for personalized experiences,
- Blockchain for secure transactions and asset/identity management,
- Computer Vision (CV) and 3D modeling for realistic environments,
- 5G/6G networks for ultra-fast connectivity,
- Edge computing for real-time processing,
- XR, VR, AR, and Mixed Reality (MR)–for immersive interaction,
- Robotics and the IoT for automation and seamless interconnectivity,
- Big Data for complex analytics,
- DTs for creating virtual replicas of real-world systems, enabling real-time simulations in urban planning, healthcare, and industry.

As the metaverse gains mainstream adoption, platforms like Meta's Horizon Workrooms and AR Calls redefine modes of remote presence, integrating gaming, social networking, and professional collaboration [16]. User engagement is facilitated through Head-Mounted Displays (HMDs), AR goggles, VR headsets, sensor networks, smartphones, and wearables [59].

Originally rooted in Hindu mythology, avatars now serve as digital identities in the metaverse, enabling users to adopt either realistic likenesses or fully personalized personas [61]. Avatars are increasingly used in professional domains such as virtual boardrooms, finance, and healthcare, as well as in social environments [64]. However, their expanding adoption, particularly in virtual offices and telemedicine, raises concerns about impersonation, fraud, and identity security, especially as younger generations increasingly gravitate toward them [71]. Against this backdrop, layered identity defenses, such as Verifiable Credentials (VCs) and Decentralized Identifiers (DIDs), combined with privacy-preserving approaches (e.g., Decentralized Privacy-Preserving Machine Learning-DPPML) are employed to mitigate risks of identity misuse, impersonation, and credential or avatar compromise without centralizing sensitive data. Accordingly, ensuring trust, inclusivity, and robust AI governance remains essential for the metaverse's sustainable evolution.

### 2.4. The emergence of CitiVerse in SCs

The evolution of SCs has shifted from a technology-driven model centered on ICT and infrastructure to a holistic, citizen-centric approach that emphasizes inclusion, sustainability, and participatory governance. This transition highlights the growing role of virtual landscapes in urban development, culminating in the concept of "CitiVerse", a metaverse-integrated SC framework designed to enhance digital interactions and urban experiences. Rather than a single platform, CitiVerse is a global metaverse ecosystem that supports open, interoperable, and innovative virtual environments. Its core mission is to tackle urban challenges while unlocking new possibilities for enhanced urban living.

CitiVerse extends metaverse principles into SCs by synchronizing city-specific virtual environments with physical spaces, fostering social trust, promoting people-centered engagement, and supporting avatar-mediated services across administrative, economic, social, cultural, and environmental domains. Defined as a subcategory of the metaverse, CitiVerse or "virtual worlds for citizens" provides new capabilities for citizens and communities, enabling active participation and augmenting real-world urban interactions. International bodies, such as the ITU, United for Smart Sustainable Cities (U4SSC), United Nations (UN), UN-Habitat, and the European Union (EU) increasingly support these trajectories to accelerate digital transformation and narrow the digital divide [72].

Operationally, CitiVerse is grounded in 3 principles that translate its vision into practice:

- CitiVerse prioritizes people-centricity. This human-centered engagement and empowerment mirrors and enhances everyday city experiences through immersive participation, transforming residents into co-designers rather than passive recipients.
- CitiVerse offers a broad portfolio of innovative virtual goods, services, and capabilities. This portfolio encompasses public, administrative, economic, social, political, cultural, and environmental services, all delivered within open and interoperable environments that ensure transparency, auditability, and reusability.
- City and community actors are represented by avatars within CitiVerse. This avatar-based representation for all city stakeholders, including citizens, civil society, and organizations, enables inclusive and multidimensional interaction across domains.

In practice, CitiVerse operates as a city-scale rehearsal environment where residents enter a high-fidelity DT as avatars and preview the consequences of interventions before physical implementation. For example, in a mobility pilot, as in Virtual Singapore [73,74], parents "walk" through the school corridor at 08:00 a.m. while buses glide through retimed signals and cyclists test protected lanes. Meanwhile, the platform logs route choices, wait times, and near-miss events, then recommends options that reduce peak-hour delays and exposure to Nitrogen Dioxide ($NO_2$) and Particulate Matter (PM). For permitting and governance, as demonstrated by Metaverse Seoul's virtual service centers [12,73], telepresence counters enable residents to file requests, attend hearings, or participate in design charrettes without commuting. Usage analytics support right-sizing office footprints and repurposing surplus parking areas into café seating, wider sidewalks, or bicycle docks. For urban greening, as seen in Helsinki 3D and Virtual Singapore climate-energy models [73–75], neighbors drag and drop trees, shade sails, and green roofs onto their block, observing real-time cooling effects on the heatmap. Scenarios are ranked by canopy gain, surface-temperature reduction, and equitable access to green space. At the parcel and street level, consistent with Helsinki 3D's Energy & Climate Atlas [73,75], homeowners simulate façade retrofits to visualize changes in solar gain and energy bills, while small businesses model pedestrian flow and delivery schedules to deconflict curb usage. During virtual town halls, similar to planning and engagement workflows on VU.CITY in London [73], participants explore 3D budget layers, seeing, for example, how reallocating 1 percent from off-street parking to pocket parks affects shade, noise, and play space. They can also query an AI civic assistant that translates urban plans into plain language. Together these capabilities make benefits tangible (e.g., shorter trips, cleaner air, fewer offices and parking stalls, more greenery) and transform participation into measurable signals that inform real-world SC decisions.

Despite extensive research on SCs and the metaverse, their integration remains largely unexplored. The metaverse presents transformative opportunities for work, governance, and social interaction, with direct implications for SCs, including new economic prospects, enhanced civic engagement, and streamlined service delivery. Additionally, significant challenges persist, notably infrastructure constraints, governance complexities, data privacy risks, and inclusivity gaps [6]. Current literature often treats SCs and the metaverse as separate domains, leaving a critical gap in understanding their joint operationalization. Addressing this gap requires targeted research on feasibility, ethical deployment, and the socio-political impact of metaverse-driven urban ecosystems.

## 3. Research methodology

This systematic literature review follows the PRISMA 2020 guidelines [76] to ensure a structured, transparent, and reproducible synthesis of studies at the intersection of the metaverse and SCs. The review adopts the 4 standard PRISMA stages, which are Identification, Screening, Eligibility, and Inclusion, with each stage documented through explicit criteria and procedures. The search strategy and selection rationale are described narratively, while **Figure 2** illustrates the PRISMA flowchart summarizing the number of records processed at each stage of the review.

The **protocol and search design (planning)** were developed to ensure methodological rigor and comprehensive coverage of relevant literature. The search strategy combined both domain-specific and application-specific terms using Boolean operators:

(metaverse OR citiverse OR "virtual world" OR "virtual worlds" OR "immersive environment" OR "immersive environments" OR "extended reality" OR "mixed reality" OR "augmented reality" OR "virtual reality" OR "digital twin" OR "digital twins" OR "twin city" OR "twin cities" OR "virtual space" OR "virtual urbanism" OR "immersive urban environment" OR "virtual governance" OR "virtual society" OR "virtual economy" OR "virtual infrastructure") **AND** ("smart city" OR "smart cities" OR "digital city" OR "digital cities" OR "intelligent city" OR "intelligent cities" OR "connected city" OR "connected cities" OR "knowledge city" OR "knowledge cities" OR "information city" OR "information cities" OR "ubiquitous city" OR "u-city" OR "cyber-city" OR "cyber cities" OR "virtual city" OR "virtual cities" OR "data-driven city" OR "AI-enabled city" OR "cognitive city" "smart sustainable city" OR "eco-city" OR "resilient city" OR "metaverse city" OR "CitiVerse" OR "metaverse-enabled city" OR "augmented city" OR "digital twin city" OR "hybrid city" OR "platform city" OR "blockchain city" OR "smart community" OR "smart region" OR "smart territory").

The search covered the period from 2018 to 2025, capturing recent advancements in metaverse applications within SC contexts. To ensure the relevance and quality of the reviewed literature, specific inclusion and exclusion criteria were defined and applied.

**Eligibility criteria:**

- Inclusion criteria included peer-reviewed journal and conference publications, doctoral dissertations, full-text availability, English language publications, and explicit relevance to the metaverse-SC domain.
- Exclusion criteria involved studies without full text access, research lacking clear connection to metaverse-SC variables, non-English publications, or duplicates.

During the **Identification stage,** an initial search across Google Scholar and Web of Science yielded 398 records, with an additional 73 items identified through complementary sources (e.g., cross-references and forward-backward searches), totaling 471 records. After automated and manual deduplication, 313 unique records remained for screening.

In the **Screening stage**, titles, abstracts, and keywords were screened for relevance against the predefined inclusion criteria (e.g., SC-metaverse context, scholarly nature, English language, and full-text availability). Non-scientific items (e.g., newsletters, editorials, policy notes) and off-topic studies were excluded. Following this stage, 208 records proceeded to full-text assessment, while 105 were excluded based on irrelevance to SC or metaverse domains.

In the **Eligibility stage,** the full texts of the remaining 208 records were examined in detail to assess methodological adequacy, analytical depth, and explicit metaverse-SC integration.

A total of 54 studies were excluded at this stage for the following reasons:

- Insufficient methodological detail (n = 12)
- Lack of analytical or empirical contribution (n = 29)
- No clear SC-metaverse linkage (n = 6)
- Non-accessible or incomplete content (n = 7)

Consequently, **154** studies met the eligibility criteria and were included in the final synthesis.

During the **Inclusion and Thematic Synthesis stage,** the **154** included studies were systematically coded using a standardized extraction template capturing: (i) bibliographic metadata, (ii) study type and geographical context, (iii) SC pillars addressed (Economy, People, Governance, Mobility, Environment, Living), (iv) metaverse capabilities (simulation, XR/immersion, co-creation/participation, data-driven services), (v) technological enablers (AI, DTs, blockchain, edge/5G, DPPML),

(vi) stakeholder groups, (vii) outcomes, and (viii) associated risks (privacy, security, inclusion, governance).

Identification
Records identified through query in scholar databases search (n = 398)
Aditional records identified through other sources (n = 73)
Records after duplicates removed (n=313)
Screening
Records screened (n = 313)
Records excluded considering the title and abstract (n = 105)
Eligibility
Full-text articles assessed for eligibility (n = 208)
Full text articles excluded due to
Reason1: 12
Reason2: 29
Reason3: 6
Reason4: 7
(n=54)
Included
Articles included in analysis (n = 154)

**Figure 2.** Flow diagram of the PRISMA methodology illustrating the results of the identification and screening stages

A thematic analysis was then conducted, organizing the literature into 3 major themes (**Table 2**), which informed the cross-city case analysis in Section 4 and the subsequent examination of challenges, risks, countermeasures, and the SWOT analysis presented in Section 5.

**Table 2.** Key themes of the review.

| No | Key-Themes |
|---|---|
| 1 | Metaverse definition and core elements |
| 2 | Metaverse and SC implementation fields |
| 3 | Challenges, risks, and countermeasures in metaverse-SCs |

This structured PRISMA-based approach ensures methodological rigor, transparency, and replicability, providing a robust foundation for analyzing the convergence of metaverse technologies and smart city systems.

## 4. Results

### 4.1. CitiVerse integration pillars in SCs for a digital future

In the evolving landscape of urban development, technology is increasingly recognized as the cornerstone of urban "smartness," not as an end in itself. This study investigates how metaverse applications influence the 6 key SC dimensions, as defined by Giffinger et al. [24], including economy, people, governance, mobility, environment, and quality of life. With nearly 700 cities projected to implement metaverse-driven initiatives by 2030 [3], understanding its impact is critical. Through a comprehensive literature review, this research identifies key themes that highlight the metaverse's role in urban evolution. The findings suggest that metaverse technologies can drive economic growth, enhance citizen engagement, streamline governance, reimagine mobility, promote environmental sustainability, and elevate overall urban living standards. The integration of the metaverse within SCs is structured around 6 key pillars (Smart Economy, Smart People, Smart Governance, Smart Mobility, Smart Environment, and Smart Living), each representing a fundamental aspect of urban transformation. These pillars highlight the diverse ways in which immersive technologies, AI, and DTs are reshaping economic models, citizen engagement, governance structures, transportation systems, sustainability efforts, and overall quality of life. The following sections explore each pillar in detail, examining both the opportunities and the challenges associated with their implementation.

#### 4.1.1. Smart Economy

The convergence of the metaverse with SCs is a transformative dynamic that redefines urban development, creating synergies that broaden economic opportunities [77]. This section explores how the metaverse reshapes core aspects of the smart economy within SCs, such as economic transformation, digital commerce, workplace evolution, brand engagement, and digital assets, while emphasizing the implications for urban environments.

**Economic transformation:** The metaverse is reconfiguring economic structures in SCs, accelerating digital transformation while challenging traditional business models. Major corporations, including Microsoft, Apple, Meta, Roblox, and Nvidia, are driving this shift by investing significantly in XR devices, immersive platforms, and digital-commerce ecosystems [6,9,16,68,78–80]. Cities, such as Dubai and Shanghai, are positioning themselves as metaverse-driven economic hubs, with Dubai targeting 40,000 jobs by 2030 and Shanghai allocating $1.5 billion to an industry projected at $52 billion by 2025 [73,81]. While these initiatives promise growth, they also pose risks of market concentration within dominant tech firms, potentially limiting accessibility [3].

**E-commerce**: In e-commerce, the metaverse blurs physical and digital transactions, enabling retailers to create immersive shopping experiences through VR and AR [3]. The shift from Direct-To-Consumer (DTC) to Direct-To-Avatar (DTA) commerce fosters hyper-personalized interactions but disrupts traditional supply chains [54,69,80]. However, lightly regulated metaverse marketplaces introduce risks, e.g., illicit trade, tax evasion, data privacy violations, and over-commercialization, raising ethical concerns over consumer manipulation, algorithmic surveillance, and exploitation [65]. SCs must balance economic growth with proportionate regulatory oversight to ensure fair trade and consumer protection.

**Workplace evolution:** Workplaces are evolving as virtual collaboration platforms, such as Microsoft Mesh, redefine remote work, facilitating global connectivity and multicultural engagement [12,82,83]. As demand grows for XR-skilled professionals, like VR/AR specialists and digital-content creators, organizations increasingly integrate DTs and IoT to enhance real-time data sharing and optimize resource allocation [16,69,84,85]. Yet this shift also introduces challenges, including blurred work-life boundaries, heightened digital-surveillance risks, and the psychological toll of prolonged virtual immersion. While avatars can enhance efficiency in training and customer service, they also raise concerns about job displacement, digital-identity ethics, and the sustainability of fully virtual workplaces [78,86,87]. Accordingly, metaverse-enabled work requires strong ethical guidelines, labor protections, and digital governance to ensure that innovation does not compromise worker well-being and equity [65].

**Brand engagement and advertising**: Brand engagement and advertising have rapidly transformed, with firms leveraging the metaverse to connect with digital-native consumers through virtual events, avatar accessories, and in-game advertising on platforms such as Roblox [62,88]. Brands, such as Coca-Cola and Volkswagen, are integrating immersive marketing strategies, with stock market trends reflecting investor confidence in the metaverse economy [29].

**Cryptocurrencies and Non-Fungible Tokens (NFTs):** Cryptocurrencies and NFTs reshape financial ecosystems in metaverse-SC contexts, enabling decentralized transactions through Bitcoin, Ethereum, and platform-specific tokens (e.g., Robux) [59,79,89,90,91]. NFT marketplaces, such as OpenSea, facilitate multi-billion-dollar trades in virtual real estate, digital fashion, and artworks, broadening digital-asset ownership. However, commercialization risks widening disparities as smaller businesses and creators struggle for visibility in platform-centric economies [84]. Additionally, reliance on digital currencies may exclude populations without financial or technological access, deepening socioeconomic divides [47]. Addressing these challenges demands inclusive financial policies that prioritize accessibility and digital equity.

#### 4.1.2. Smart People

The metaverse is redefining SCs by fostering citizen engagement, participatory governance, and broader digital inclusivity. The following key aspects illustrate how metaverse contributes to the development of smart citizens.

**Urban planning:** The metaverse in SCs enhances urban planning through advanced modeling and simulation, allowing residents to visualize and co-design people-centered cityscapes, thereby strengthening civic ownership and engagement [45,88]. Initiatives such as the Liberland Metaverse [92] and Metaverse Seoul [73] demonstrate how DTs, AR, and immersive platforms empower citizens to shape urban environments, optimize spatial performance, and expand democratic decision-making [11,12,68,93].

**Interactive learning experiences:** Education in the metaverse transcends classroom boundaries, fostering interactive, immersive learning that supports lifelong education and vocational training [62,94]. Universities such as Stanford, Embry-Riddle, and Case Western Reserve have integrated VR-based courses, enabling students to attend classes in simulated environments ranging from museums and emergency-aviation response to anatomical labs [15,95]. Vocational programs, including Seoul Citizens' University Metaverse Campus and Seoul Run, equip professionals in healthcare, fashion, and emerging industries, while Zhejiang University's "Waste to Energy" virtual workshop enabled global collaboration during the COVID-19 pandemic [85,96]. However, metaverse-based education requires robust digital citizenship norms to prevent misinformation, cyberbullying, unethical conduct, and identity deception. While it can foster self-expression through avatars and role-playing, it may blur social norms and amplify biases related to gender, ethnicity, and socioeconomic status, underscoring the need for clear governance and responsible-use policies [14].

**Citizen engagement and inclusivity:** Beyond education, the metaverse enhances inclusivity in SCs through AI-enabled multilingual tools that reduce linguistic barriers [59]. Neural Machine Translation (NMT) and Natural Language Processing (NLP) enable real-time adaptation to diverse languages, dialects, and cultural nuances, broadening equitable access to public services and digital governance [14,16,97–99]. Drawing on Arnstein's Ladder of Citizen Participation [100], metaverse applications introduce new democratic models by enabling residents to engage in co-design, crisis response simulations, and smart policy formation. Large-scale translation models, such as Meta's No Language Left Behind (NLLB-200), facilitate cross-linguistic participation in governance, supporting digital town halls, e-voting systems, and community-driven initiatives. Additionally, VR-based co-simulation enables real-time collaboration between policymakers and citizens [62], [88,101], strengthening participatory governance [40,102,103]. Yet this promise comes with risk; the metaverse can deepen digital inequalities, as factors such as socioeconomic status, education, age, and disability may hinder access, thereby reinforcing systemic disparities in technology adoption. This "metaverse divide" could benefit the tech-savvy while marginalizing underserved populations, exacerbating stratification rather than bridging it [16]. Moreover, technical feasibility does not guarantee public acceptance, which highlights the need for inclusive policies, accessibility by design, device support, and digital-skills programs to avoid widening existing social divides.

#### 4.1.3. Smart Governance

The integration of the metaverse into SCs is reshaping governance by fostering greater inclusivity, transparency, and participatory decision-making [45,104]. Immersive digital platforms enable governments to enhance citizen engagement in policymaking, service delivery, and crisis management, aligning governance with sustainable urban development.

**Inclusive and participatory governance:** The metaverse acts as a catalyst for inclusive governance, offering e-democracy tools that facilitate real-time, two-way digital participation [3]. Platforms, such as Metaverse Seoul and metaverso.navarra.es, provide virtual spaces where citizens interact with policymakers, co-design infrastructure projects, and access digital services [12,68,105]. However, the digital divide remains a pressing challenge, as disparities in access to high-speed internet and advanced devices can marginalize lower-income communities, undermining equitable governance [46].

**Political engagement and digital campaigns:** The metaverse is also transforming political participation by expanding interactive digital campaigning and voter outreach. The Biden-Harris (2020) campaign leveraged the virtual world of "Animal Crossing: New Horizons" to engage younger, tech-savvy voters, while Congresswoman Alexandria Ocasio-Cortez conducted virtual "house-to-house" (island-to-island) visits to connect with constituents during COVID-19 restrictions [105]. These strategies highlight the metaverse's potential to reach digital-native voters, yet concerns remain regarding algorithmic amplification, misinformation, and the exclusion of non-tech-savvy demographics [59].

**Citizen co-creation and feedback mechanisms:** Metaverse applications strengthen citizen co-creation, allowing residents to actively contribute to urban planning and policy formation [6]. Digital feedback loops provide real-time insights, enabling governments to iterate policies based on citizen input [3,85,103]. This participatory model fosters transparency, builds trust, and ensures governance remains responsive to community needs.

**Virtual service delivery and crisis management:** Metaverse technologies are transforming remote service delivery, enabling access to public services, healthcare, and education without physical presence [12]. Cities, such as Shanghai, Singapore, and Helsinki, leverage DTs to optimize resource allocation and improve accessibility and efficiency in urban services [68,73], while Barbados has partnered with Decentraland to establish virtual embassies and consulates, offering e-visas and remote diplomatic services [73,105]. Additionally, metaverse-based crisis simulations enhance disaster preparedness, providing citizens with interactive, scenario-based emergency response training [106,107]. However, the immersive nature of the metaverse also introduces risks of manipulation and misinformation, potentially biasing decision-making, distorting public discourse, and eroding trust in governance [3].

**Transparent governance and virtual town halls:** Transparency and accountability are reinforced through metaverse-driven open government. Virtual town halls and interactive citizen assemblies provide immersive policy discussions, ensuring greater public accountability [45,46,101]. Open-data platforms combat corruption by offering real-time tracking of urban projects. However, concerns over privacy and cybersecurity persist. The potential for unauthorized surveillance and data exploitation underscores the need for clear, enforceable regulatory frameworks to balance innovation with ethical governance [59]. As SCs continue to integrate metaverse technologies, the challenge lies in ensuring these innovations serve as tools for democratic enhancement rather than instruments of exclusion or manipulation. Establishing robust digital-literacy programs, privacy-preserving safeguards, and accessibility-by-design policies will be critical to fostering a governance model that is truly inclusive, transparent, and citizen-centered.

#### 4.1.4. Smart Mobility

The adoption of metaverse technologies in SCs is reshaping mobility, transportation, navigation, and tourism by enhancing sustainability, efficiency, and inclusivity. Leveraging AR, AI-driven simulations, and real-time data, SCs optimize public transport, improve safety, and provide immersive travel experiences [9].

**Smart navigation and productivity in urban spaces:** Metaverse-powered AR navigation tools improve mobility within indoor and public spaces, where a large share (90 percent) of daily activities occurs [2]. Interactive maps, real-time route optimization, and accessibility overlay streamline movement across smart campuses [108], business districts, hospitals, malls, and airports [13,109]. Workplace productivity also benefits from AR-guided navigation integrated with building information modeling, optimizing efficiency in construction sites, industrial zones, and logistics hubs [9,110]. However, continuous pedestrian tracking raises concerns about data surveillance and misuse of personal location data [65]. While high-tech SCs, like Singapore and Shanghai, leverage DTs for real-time movement insights and infrastructure monitoring [73], smaller cities with limited capacity risk falling behind, thus exacerbating the digital divide [46].

**Smart transportation systems and Autonomous Vehicles (AVs):** As cities evolve into digital SCs, the metaverse supports smart transportation through AI-based traffic management, vehicle-to-vehicle (V2V) communication, and predictive analytics, thereby reducing congestion and boosting efficiency [111,112]. AVs benefit from metaverse-based route simulations that enhance navigation, safety, and fleet management [113]. Real-time transport apps assist citizens by providing AI-informed insights on schedules, alternative routes, and congestion levels, helping to reduce $CO_2$ emissions and optimize travel [78,93]. However, the rise of AVs raises challenges related to liability in accidents and the potential displacement of drivers and transportation workers [56].

**Enhanced safety and emergency response:** The metaverse also enhances safety through real-time pedestrian-flow tracking, optimized emergency response systems, and AI-enabled hazard detection [111,114]. AR interfaces in transportation provide predictive, context-aware guidance, reducing accident risks [107]. In the maritime domain, AR-based navigation improves surface and subsea mobility, while Remotely Piloted Systems (RPS) and Unmanned Aerial Vehicles (UAVs) benefit from metaverse-enabled aerial navigation [111,112]. SCs, such as Dubai and Seoul, use virtual emergency simulations to train first responders, refine crisis management, and strengthen urban resilience [73,107].

**Immersive smart tourism and cultural accessibility**: Metaverse-driven tourism is redefining cultural accessibility in SCs through AR-enhanced city tours, digital storytelling, and virtual heritage preservation [3,107]. SCs, like London and Helsinki, offer immersive museum experiences and interactive 3D exhibitions, deepening cultural engagement [95,107]. Virtual tourism platforms can make iconic locations accessible to individuals with disabilities or travel restrictions, thereby advancing inclusion. However, the high cost of VR equipment may exclude lower-income individuals from premium metaverse experiences, widening disparities. Additionally, virtual tourism may risk diminishing in-person visits and affecting local economies reliant on traditional tourism [3]. As metaverse technologies reshape mobility and tourism, balancing innovation with equity and ethical governance remains crucial. Ensuring broad accessibility, safeguarding data privacy, and developing inclusive policies will be essential for realizing the full potential of metaverse-supported urban mobility.

#### 4.1.5. Smart Environment

Incorporating metaverse technologies into SCs offers transformative solutions for environmental sustainability, including reducing carbon footprints, optimizing resource management, and fostering climate resilience. Through DTs, AI-driven simulations, and immersive collaboration tools, SCs advance sustainable urban planning, enhance crisis management, and promote community-driven environmental initiatives [115]. However, while these innovations hold promise, they also raise concerns about energy consumption, digital accessibility, and socio-environmental equity.

**Reducing carbon footprints and supporting global climate initiatives:** The metaverse supports global climate initiatives, such as the Paris Agreement and the Sustainable Development Goals (SDGs), by helping to reduce transportation emissions and other energy-intensive operations [3,16]. Virtual collaboration platforms, such as Microsoft Mesh, enable professionals to conduct meetings in digital spaces, thereby minimizing travel and office energy consumption [67]. Tech giants, like Meta and Microsoft, are advancing sustainability efforts, committing to net-zero emissions and integrating renewable energy into their digital infrastructure by 2030 [116]. SCs, such as Shanghai, Singapore, and Helsinki, deploy metaverse-powered DTs to monitor carbon footprints, optimize energy consumption, and implement climate adaptation strategies [73]. However, the sustainability benefits of metaverse adoption can be offset by high computational demands; data centers, cloud storage, and AI-driven simulations require substantial energy, which-if reliant on fossil fuels-may undermine emissions reductions [3].

**Smart resource management and sustainable urban design:** Smart resource management in SCs is enhanced through metaverse-enabled digitalization, reducing reliance on physical infrastructure and promoting more sustainable consumption patterns [115]. SCs, such as Shanghai, Singapore, and Helsinki, leverage metaverse-integrated DTs for real-time urban monitoring, supporting efficient energy distribution and informing climate adaptation measures [73]. The metaverse can optimize public transportation by nudging shifts from private vehicles to low-carbon transit systems, thereby reducing urban congestion and carbon emissions. However, disparities in technological capacity and access may impede equitable uptake, potentially exacerbating the divide between well-equipped SCs and under-resourced urban areas [46].

**Urban resilience and environmental crisis management:** Urban resilience is strengthened through metaverse-integrated DTs that enhance environmental crisis management and disaster preparedness [115]. Predictive models, such as those used in Boston's GoBoston 2030 initiative, simulate climate risks (e.g., flooding and wildfires), improving response strategies [67,116]. SCs, like Dubai and Seoul, deploy metaverse-based emergency-training simulations, equipping first responders with real-time crisis-response techniques and bolstering urban resilience [107].

**Community engagement and sustainable living:** Community-driven sustainability initiatives thrive in metaverse-enabled SCs, where citizens actively participate in projects such as urban gardens and climate awareness campaigns [117]. Platforms in Singapore and Helsinki enable residents to co-design eco-friendly infrastructure, expanding public involvement in sustainability [107]. Additionally, metaverse-enhanced remote work can alleviate urban congestion, support thoughtful deconcentration, and reduce environmental strain on high-density areas [3,116]. Nevertheless, these opportunities are not universally accessible; marginalized groups and low-income communities may lack the necessary technology and internet access, limiting participation and reinforcing environmental inequities [46]. Overall, while the metaverse provides a promising framework for SC sustainability, success depends on equitable access, clear regulatory oversight, and integration with sustainable energy systems. Addressing digital divides and ensuring that environmental benefits reach all urban populations will be critical to building a greener, more resilient future.

#### 4.1.6. Smart Living

The metaverse is reshaping the social dimensions of SCs by enhancing self-expression, entertainment, education, healthcare, and cultural preservation. It fosters inclusivity, creativity, and well-being by transcending physical and geographical barriers, democratizing knowledge, and improving healthcare accessibility, ultimately redefining human interactions within smart urban ecosystems [118]. However, these benefits bring ethical, social, and technical challenges, necessitating regulatory frameworks that ensure equitable access, strong data security, and sustainable integration of virtual experiences into SCs.

**Enhanced social interaction, self-expression, and entertainment:** SCs increasingly leverage metaverse platforms to create dynamic social environments that promote digital inclusion and real-time connectivity. Cities, such as Seoul [52] and New York City [119], use virtual spaces for public events, digital gatherings, and cultural experiences, broadening access to urban life. Museums and galleries in London and Helsinki integrate metaverse-enhanced exhibitions, allowing citizens to engage with art and history [3,95]. Social platforms, like VRChat and Rec Room, enable avatar-based interaction, strengthening communal ties and fostering digital self-expression [117]. However, these innovations may also contribute to social isolation, digital addiction, and erosion of in-person interaction, underscoring the need for balanced, time-aware virtual-engagement policies [65].

**Education, digital learning, and knowledge accessibility:** SCs incorporate metaverse technologies to deliver immersive, interactive learning experiences. Platforms, like Singapore's Virtual Academy and Helsinki's digital-education initiatives, align with the SDGs by improving access to quality education and vocational training [96,99,116]. Metaverse-driven simulations in healthcare, engineering, and aviation provide hands-on training, preparing professionals with job-relevant skills [62,94]. Virtual libraries, such as the Community Virtual Library (CVL), democratize knowledge via 24/7 access to academic resources, reducing traditional barriers [109]. Nevertheless, educational inequalities persist; marginalized communities may lack required devices and connectivity, while overreliance on virtual learning can dilute essential hands-on practice, thus supporting hybrid models to preserve skill development [120].

**Healthcare and well-being:** The metaverse enhances public-health services in SCs through telemedicine, AI-driven diagnostics, and virtual-therapy spaces [19], [121]. Cities such as Dubai [122] and Singapore [74] implement remote surgery, real-time patient monitoring, and mental-health support, improving access and efficiency in medical care [121,123]. Additionally, AR-based fitness programs and metaverse-integrated wellness initiatives encourage healthy urban lifestyles in smart communities [3,124]. However, privacy and cybersecurity concerns-along with the risk of excluding older or less tech-savvy users-remain salient; AR headsets capturing biometric/spatial signals require accessibility-by-design and privacy-preserving analytics (e.g., DPPML) [14,121,125,126].

**Cultural heritage preservation and inclusive tourism:** Cultural heritage preservation is increasingly facilitated through metaverse technologies, supporting long-term accessibility to historical artifacts and landmarks [12]. Digital archives, 3D-rendered heritage sites, and interactive AR museum tours extend cultural engagement to broader audiences, including individuals with disabilities or geographic constraints [3,93,95]. SCs, such as London and Helsinki, leverage these tools to promote inclusive tourism, reinforcing the metaverse's role in global cultural accessibility and heritage conservation. Yet, virtual heritage initiatives can risk over-commercialization and diminished authenticity; balancing digital access with on-site immersion is essential to preserve historical integrity [68,111,114].

### 4.2. Real-world applications: Case studies across metaverse-SC domains

The integration of the metaverse within SCs unlocks transformative advancements across governance, mobility, economy, sustainability, and quality of life for smart citizens. By leveraging AI, DTs, and XR (VR/AR), SCs promote transparency, enable participatory governance, and expand access to public services, exemplified by initiatives like Metaverse Seoul and Singapore's digital-government platforms. In mobility, real-time traffic management, AV simulations, and smart navigation systems increase efficiency and reduce congestion. The metaverse drives economic innovation by supporting digital commerce, workplace evolution, and decentralized financial ecosystems, attracting global investments, as seen in Dubai and Shanghai's metaverse strategies. Sustainability efforts benefit from metaverse-enabled climate modeling, collaborative planning, and resource optimization, lowering carbon footprints and strengthening urban resilience. Education and healthcare are enhanced through immersive learning environments, AI-driven vocational training, telemedicine, and remote diagnostics, improving accessibility and efficiency. Additionally, cultural-heritage preservation and inclusive tourism-such as metaverse-powered museum experiences in London and Helsinki- broaden access to historical sites. Collectively, these deployments demonstrate how integrating metaverse technologies helps SCs build more inclusive, efficient, and sustainable urban ecosystems, redefining the future of smart living.

**Table 3** presents a comparative analysis of metaverse applications in SCs across various domains. The cities included in this analysis were selected based on their prominence in literature and their ongoing initiatives related to metaverse integration. These cities are referenced in academic studies, policy reports, and industry analyses, underscoring their pioneering role in adopting metaverse-driven solutions. The selection also reflects regional diversity, covering SCs from Asia, Europe, North America, and the Caribbean, each demonstrating unique approaches to urban governance, economic development, sustainability, mobility, and digital inclusion.

**Table 3.** Case studies of metaverse applications in SCs

| SC | Metaverse applications in SCs |
| --- | --- |
| Seoul (South Korea) | Seoul is at the forefront of metaverse-driven smart urbanism with the Metaverse Seoul project, launched under the Seoul Vision 2030 initiative. This three-phase program (2021–2026) integrates AI, blockchain, and XR to enhance digital governance, economic participation, and civic engagement. The city enables residents to handle administrative tasks, taxes, and legal matters through virtual platforms, improving accessibility and efficiency. Additionally, Metaverse Seoul fosters innovation by supporting startups and creating digital workspaces. Yet persistent risks, like digital exclusion among lower-income and elderly residents with limited access to VR devices and high-speed connectivity, alongside privacy and security concerns, and the energy footprint of AI/blockchain components, remain material challenges that require targeted inclusion programs and robust governance. Seoul's approach is groundbreaking, but its long-term success depends on balancing inclusivity, security, and sustainability [12,73]. More information can be found on the Official Website of the Seoul Metropolitan Government (https://english.seoul.go.kr) |
| Dubai (United Arab Emirates) | Dubai's Metaverse Strategy, launched by H.H. Sheikh Mohammed bin Rashid Al Maktoum, aims to position the city among the top 10 global metaverse economies. By 2030, the initiative plans to attract over 1,000 blockchain and metaverse companies, create 40,000+ jobs, and drive economic growth. The government is advancing AI-driven governance, Web3 innovation, and immersive digital services to enhance investment opportunities and smart infrastructure. Virtual real estate transactions and AI-powered healthcare solutions, including telemedicine and smart hospitals, demonstrate Dubai's commitment to applying metaverse technologies across multiple sectors. However, digital inequality remains a concern, as access to these innovations often depends on high-end technology and digital literacy. Additionally, cybersecurity threats, financial speculation in virtual real estate, and regulatory uncertainties pose risks to long-term stability. Dubai's aggressive push into the metaverse is transformative but requires strong regulatory frameworks and inclusive digital policies [56,81]. Official information and updates are available on the Dubai Government Portal (https://u.ae/en/#/). |
| Tampere (Finland) | Tampere's Metaverse Vision 2040 outlines a structured roadmap for integrating AI, DTs, and immersive technologies into urban infrastructure. Through its CitiVerse initiative, the city collaborates with European partners, such as Rotterdam and Istanbul, to develop an ethical and human-centric metaverse ecosystem. Tampere is testing AI-powered planning tools, predictive modeling, and metaverse-enabled education to support a sustainable and inclusive SC transition. However, the city faces significant challenges, including high costs, potential digital inequality, and privacy concerns related to AI governance and data collection. Moreover, |

Continuation Table:

| SC | Metaverse applications in SCs |
|---|---|
| | despite the goal of becoming carbon-negative, the high energy demand of metaverse infrastructure could undermine sustainability efforts. Tampere's model is innovative, but balancing technology, accessibility, and environmental responsibility will be critical [73,75]. Additional details are available on Metaverse Institute's official page (https://metaverse-institute.org/). |
| Helsinki (Finland) | Virtual Helsinki, developed by ZOAN in 2018, is a highly detailed DT that enables virtual tourism, cultural events, and immersive retail experiences. Unlike many newer metaverse initiatives, Helsinki implemented VR-based experiences before the 2021 metaverse wave. The platform provides interactive tours of landmarks, concerts, and exhibitions, positioning the city as a leader in smart tourism and digital governance. However, awareness and accessibility issues persist, as not all citizens have the technology required to access Virtual Helsinki. While the project meets metaverse criteria, it lacks branding associated with more recent initiatives. Furthermore, socioeconomic barriers and technological access must be addressed to ensure broader participation. Helsinki's proactive approach to digital urbanism is commendable, but its full potential depends on greater inclusivity and public engagement [73,75].<br>Further information about Helsinki's Virtual City and 3D DT initiatives can be found on ZOAN's official website (https://zoan.com/). |
| Catalonia (Spain) | Catalonia's CatVers metaverse, developed by the Catalan Blockchain Centre (CBCat) with support from the Barcelona Chamber of Commerce, focuses on preserving Catalan identity, fostering decentralized governance, and creating a regional digital space. CatVers hosts virtual meetings, art exhibitions, and cultural festivals, reinforcing Catalonia's commitment to digital sovereignty and innovation. Plans for metaverse university campuses further highlight its educational and research potential. However, the primary use of the Catalan language can restrict access for non-Catalan speakers, limiting global reach and inclusivity. Additionally, economic sustainability concerns arise as the metaverse remains a speculative industry with uncertain long-term adoption. While CatVers strengthens cultural heritage through digital means, its success will depend on balancing regional identity with broader accessibility and engagement.<br>Further details can be found on the official CatVers portal (https://www.catvers.cat/). |
| London (United Kingdom) | London integrates advanced 3D modeling, AI, and AR into urban planning, improving construction efficiency, public engagement, and city governance. The VU. CITY platform, based on DT, covers 1,619 square kilometers, allows authorities to assess building density, environmental impact, and infrastructure development, improving planning transparency, and sustainability. The integration of AR-based tourism experiences and digital reconstructions of historical sites strengthen London's cultural and technological leadership. However, the focus on corporate and government applications over broader citizen-driven engagement can limit public involvement. At the same time, expanding AI-enabled public-space analytics raises privacy and ethical concerns, particularly regarding scope, purpose, and oversight of data collection and monitoring. London's digital urban transformation is impressive, but ensuring inclusivity, transparency, and ethical governance is essential for long-term success [73].<br>Further information is available on VU.CITY's official website (https://www.vu.city/cities/london). |
| New York City (United States) | New York is integrating the metaverse into urban infrastructure through projects such as Columbia University's DT initiative, which uses AI and real-time sensor data to optimize traffic flow, reduce congestion, and lower emissions. The city's cultural and commercial sectors also leverage metaverse technologies, with institutions like the Metropolitan Museum of Art (the MET) adopting virtual exhibitions, while New York Fashion Week incorporates NFTs and digital fashion experiences. |

Continuation Table:

| SC | Metaverse applications in SCs |
|---|---|
| | However, the capitalist focus on investment opportunities rather than social integration has raised concerns over corporate monopolization, insufficient government oversight, uncertain third-party access, and privacy risks. The emphasis on crypto-based commerce further highlights the speculative nature of the metaverse economy. Despite technological advancements, New York's metaverse strategy needs stronger regulation and public engagement to prevent Big Tech dominance and data exploitation [73,127]. A central question remains, "will the city leverage innovation for all, or will digital transformation deepen existing inequalities?" [119]. Further information about New York City's digital transformation and smart governance initiatives can be found on the NYC Office of Technology and Innovation website (https://www.nyc.gov/oti). |
| Shanghai (China) | Shanghai's 5-Year Metaverse Development Plan aims to create a $52 billion metaverse industry by 2025, investing $1.5 billion in AI, blockchain, VR, and cloud computing. The city plans to establish 2 industrial parks in Zhangjiang and Caohejing to foster technological innovation and accelerate digital transformation. The metaverse is expected to enhance corporate efficiency, enable new business models, and strengthen Shanghai's global tech leadership. However, concerns over government control, potential surveillance, and limited decentralization could hinder user freedom and data security. While Shanghai's metaverse economy is poised for massive growth, maintaining transparency, regulatory balance, and digital rights protections will be crucial [73]. Further information is available on the official Shanghai Government English portal (https://english.shanghai.gov.cn/). |
| Singapore (Southeast Asia) | Virtual Singapore, launched in 2014, is one of the most advanced DT platforms, integrating geospatial, demographic, and environmental data for urban planning, disaster management, and infrastructure optimization. The city also utilizes metaverse-supported healthcare (AI diagnostics, telemedicine, real-time monitoring) to improve medical accessibility. However, centralized data-management models raise privacy and data-sovereignty concerns, and the high cost of digital infrastructure may exclude lower-income citizens. Singapore's metaverse approach is highly advanced and government-driven but ensuring equitable access and robust cybersecurity remains essential [73,74]. Detailed information about the Virtual Singapore initiative can be found on the National Research Foundation's official website (https://www.nrf.gov.sg/). |
| Barbados | Barbados became the first nation to establish a virtual embassy in the metaverse (2021) through a Decentraland partnership, marking a groundbreaking move in digital diplomacy and global outreach. This initiative, led by the Ministry of Foreign Affairs and Foreign Trade, aims to promote tourism, attract investment, and expand diplomatic services. Barbados has also secured agreements with Somnium Space and SuperWorld, enabling virtual embassies and e-visa services, thus enhancing accessibility and efficiency. Additionally, the concept of a "teleporter" for cross-metaverse interoperability positions Barbados at the forefront of metaverse-driven global engagement. However, challenges include security risks, regulatory uncertainty, and jurisdictional limitations in digital diplomacy. The reliance on blockchain and decentralized platforms also exposes the initiative to market volatility and technological dependency. While Barbados is pioneering digital diplomacy, sustained success will require regulatory stability and strong cybersecurity [73,105]. More information is available on the official website of the Ministry of Foreign Affairs and Foreign Trade of Barbados (https://www.foreign.gov.bb/). |
| Boston (United States) | Boston's GoBoston 2030 and eHealth Plan integrate AI, Big Data, and IoT to enhance urban mobility and digital healthcare services. The GoBoston 2030 initiative focuses on zero-fatality, zero-injury, zero-carbon-emission transportation, using AI-driven traffic management, |

Continuation Table:

| SC | Metaverse applications in SCs |
|---|---|
| | and predictive analytics to optimize public transport and road safety. Meanwhile, the eHealth Plan (Massachusetts Digital Health Initiative) promotes telemedicine, AI-driven diagnostics, and cloud-based healthcare solutions, improving patient care while reducing costs. However, concerns persist regarding technological accessibility, potential AI bias in planning and healthcare decisions, and data privacy risks. Additionally, the economic feasibility of large-scale AI integration poses sustainability challenges. While Boston is at the forefront of AI-powered urban governance, ethical AI, inclusivity, and affordability will be critical for long-term success [73,128].<br>Details and policy updates can be accessed via the City of Boston's official GoBoston 2030 page(https://www.boston.gov/departments/transportation/go-boston-2030). |

## 5. Discussion

### 5.1. General observations from literature review

The implementation of a sustainable and accessible metaverse within SCs faces multifaceted challenges that must be addressed to ensure an inclusive and equitable digital transition.

**Technical and economic constraints:** Metaverse infrastructure demands high-performance computing, advanced networking, and vast storage capabilities, resulting in capital and operational expenditures. The high cost of XR technologies, platforms, and immersive tools-such as VR headsets-often exceeds that of traditional communication or educational media, raising concerns about exacerbated economic inequalities and financial constraints across SC projects, which may jeopardize long-term viability [14,15,129,130]. Moreover, embedding metaverse functionalities into governance requires significant public investment. Uneven resource allocation may deepen existing social and spatial inequalities, leading to exclusionary, data-driven decision-making that fails to represent diverse urban populations [3,131,132]. City-level examples illustrate these cost-related disparities in practice: Seoul's "Metaverse Seoul" is publicly financed [3,73], Singapore's "Virtual Singapore" channels significant national funding into a high-fidelity DT program [73], and London's 3D planning model (VU.CITY) operates as a commercial platform, increasing access yet reflecting private-sector priorities [73]. Similarly, Shanghai's metaverse roadmap also depends on public investment in AI/VR and cloud infrastructure [73]. Although the mix of public and private actors varies, the associated costs remain considerable. The dependency on advanced XR stacks and devices further reinforces disparities in affordability and technological capacity [3,129,132,133].

**Accessibility and digital inclusion:** Usability and access barriers remain pronounced. Vulnerable groups-such as low-income citizens, persons with disabilities (especially blind or deaf users), older adults, and non-tech-savvy users-face significant challenges in navigating immersive environments, potentially reinforcing exclusion rather than alleviating it [3,14,46,129–133]. In addition to these structural constraints, citizen mindsets resistant to digital transformation further hinder uptake and engagement, particularly among older or digitally hesitant populations [3,129]. Several city deployments make these limitations visible: Helsinki's "Virtual Helsinki" promotes cultural participation and remote urban exploration but highlights hardware cost barriers [73]. London's 3D planning model presumes baseline digital skills and compatible devices [73]. Seoul's virtual administrative services require stable connectivity and device ownership for meaningful engagement [73]. Catalonia's "CatVers" platform, though civic-oriented, demonstrates how linguistic scoping can limit access for non-Catalanian speakers [73]. Collectively, these examples underline the need for accessibility by design, while also revealing that affordability, skills gaps, and interface inclusivity remain critical bottlenecks [3,129,131].

**Ethical and behavioral considerations:** Immersive, sensor-rich environments raise significant ethical concerns, particularly around privacy, data protection, and user autonomy [3,90,93,130,132,133]. The extensive collection of behavioral and biometric data fuels anxieties about surveillance capitalism and weakens the legitimacy of user consent, posing serious governance challenges [65,133,134]. Additionally, overreliance on virtual interactions can erode physical social bonds, intensify digital escapism, and trigger mental-health concerns, especially when real-world relationships are substituted by algorithmically mediated environments [135]. Practical deployments expose these tensions: Seoul's and Singapore's metaverse strategies operate large-scale, data-intensive platforms with layered DT functionalities, where data minimization, access control, and transparent consent mechanisms remain under scrutiny [73,132]. London's AR-enhanced planning tools enable participatory design yet require stringent safeguards for public data [73,132]. Dubai's sensor-enabled transport twin and New York City's traffic-DT system similarly raise questions about data proportionality, accountability, and the risk of function-creep in civic systems [73,132].

**Legal ambiguities and governance gaps:** The legal landscape governing metaverse environments remains fragmented and underdeveloped, generating uncertainty over jurisdiction, digital assets, intellectual property, and cross-border data flows [14,134]. Disputes may arise over the regulation of cryptocurrencies, NFTs, and online advertising, as well as the role of data intermediaries and the legal status of metaverse-native entities [3,134]. Furthermore, platform centralization by dominant tech firms raises competition and labor-rights concerns. Control over virtual land, marketplaces, and transaction layers can stifle smaller actors and inflate participation costs. To prevent monopolistic tendencies, regulatory frameworks must enforce fairness, economic inclusivity, and open access to digital ecosystems [131,134,135]. However, regulatory approaches diverge across jurisdictions. In the EU, the General Data Protection Regulation (GDPR) and mandatory Data Protection Impact Assessments (DPIAs) set binding obligations for immersive systems [134]. However, city-level initiatives reflect legal pluralism rather than converge: Seoul, Singapore, Dubai, Helsinki, and London each follow context-specific frameworks [73]. Barbados' virtual embassy further complicates sovereignty and legal interoperability by extending state functions into platform-controlled spaces [73,134]. While increasing legal clarity could support institutional trust, it also heightens compliance complexity, especially across international deployments [134].

**Infrastructure, literacy, and capacity gaps:** Robust infrastructure is a prerequisite for scalable metaverse adoption. In many regions, limitations in last-mile connectivity, low bandwidth, technology discontinuation, or unstable networks constrain participation and inhibit real-time immersive experiences [44,45,47,129]. In parallel, digital literacy remains uneven, as many citizens lack the skills needed to engage effectively with extended reality systems, thus reinforcing the digital divide [46,101,129,131,133,134]. Pilot deployments across Helsinki, London, Seoul, and New York confirm that metaverse systems depend on high-speed connectivity, advanced devices, and continuous user support. Without parallel investments in affordable access and digital upskilling, their civic impact remains constrained [73,129].

**Trust deficits and public perception:** Citizen trust is a pivotal factor in the uptake of metaverse technologies in governance. Concerns about surveillance, data misuse, and opaque algorithmic systems may suppress engagement and erode legitimacy [3,130,132]. Field evidence reflects this tension: Seoul's phased-rollout strategy enhances public awareness but depends on continuous transparency about data practices [73]. London's planning tools promote trust by allowing residents to interrogate design proposals [73]. In Singapore, service efficiency must be carefully balanced against privacy concerns [73,130]. Catalonia's CatVers also shows how language-restricted interfaces can undermine inclusivity and alienate portions of the public [73].

**Cybersecurity and data sovereignty**: The convergence of XR, AI, and DT infrastructures-along with disorganized data management-significantly expands the cybersecurity threat landscape in metaverse-SCs [129,132,135]. Immersive devices used in public spaces capture multimodal data-including Red-Green-Blue-plus-Depth (RGB-D) video, audio, inertial signals, and eye-tracking-raising concerns about privacy, data security, and digital sovereignty [135]. Four key risk domains emerge from these capabilities: (a) bystander privacy violations, resulting from inadvertent recording of individuals in public or semi-private spaces, (b) Simultaneous Localization and Mapping (SLAM) leakage, which may expose sensitive physical layouts or critical city assets, (c) biometric inference and re-identification based on behavioral patterns such as gaze, gait, and motion, and (d) cross-border data processing, where sensitive urban data is handled by third-party platforms beyond the jurisdictional reach of local regulations [129,134]. These risks are compounded by uneven last-mile connectivity, lack of edge security, and dependency on opaque commercial infrastructures. In parallel, urban-metaverse systems face evolving cyberthreats, including: (i) deepfake-enabled impersonation and avatar hijacking, facilitated by Generative Adversarial Network (GANs), which can lead to identify theft and social-engineering attacks [136–138]; (ii) adversarial Machine Learning (ML) and sensor spoofing, targeting the perception stacks of XR systems to manipulate inputs and behavior [137]; (iii) telemetry-stream hijacking and replay attacks compromising the integrity of immersive experiences [137,138]; (iv) bot and Sybil attacks in digital town halls, distorting deliberative processes and undermining the legitimacy of e-participation [137]; and (v) jurisdictional ambiguity and data-sovereignty gaps, especially when city datasets are stored or processed by external providers under divergent legal obligations [137,139]. Moreover, the absence of robust backup and recovery strategies exposes metaverse systems to data loss and prolonged service disruption in the event of technical failures [132]. The academic literature on SC and immersive governance highlights these attack vectors-emphasizing threats such as gaze and gait leakage, telemetry spoofing, deepfakes, and unregulated cross-border flows [129,132,134]. Operational realities mirror these concerns: Seoul and Shanghai industrial and port-DT deployments have expanded attack surfaces across distributed edge nodes. Singapore's heavy reliance on third-party cloud infrastructure raises unresolved questions about cross-border data governance. In London, digital-engagement platforms face increased exposure to bot-based manipulation due to insufficient user-authentication measures. Despite differences in national regulatory regimes, recurring vulnerabilities-such as deepfake abuse, telemetry replay, and SLAM leakage-persist across deployments. These underscore the urgent need for privacy-by-design architectures, auditable data lineage, and zero-trust cybersecurity protocols tailored to immersive urban environments [129,132,134].

Cybersecurity is no longer optional in SCs-it is a strategic requirement to support safe digital transformation and civic engagement [135]. Countermeasures span identity, content integrity, network/edge security, analytics, governance, and operations and explicitly align with the previously identified challenges and risks. To restore trust and reduce identity/credential/avatar theft in e-participation, cities implement privacy-preserving authentication by deploying DIDs and VCs with selective disclosure, adopting passkey-based phishing-resistant logins, and using instant, continuous, on-device biometric liveness [137,138]. To curb misinformation and official/citizen impersonation via deepfakes, they reinforce content and actor integrity through provenance controls and run deepfake-detection pipelines tuned for GAN-generated media [137,139]. To contain session hijacking, telemetry replay, and adversarial sensor abuse at the edge, they adopt zero-trust access, enforce end-to-end encryption, and maintain strong device posture-hardware roots of trust, signed firmware, remote attestation of XR endpoints, and centrally managed XR fleets [136,139]. To protect spatial privacy, addressing SLAM/map leakage, bystander exposure, and continuous location/behavior tracking, they apply SLAM minimization (on-device mapping, map redaction, room-scale-only retention) and transmit pseudonymous telemetry. To analyze sensitive civic/DT data while respecting data sovereignty and ensuring legal compliance, they combine DPPML (federated/peer-to-peer training with secure aggregation and differential privacy) with confidential computing for high-risk workloads [125,126,136,139]. On governance, to resolve legal ambiguity and compliance burdens-especially under GDPR-cities conduct DPIAs for cross-border processing (these are required), codify data-sovereignty rules (localization, purpose limitation, auditable lineage), and bind third-party/cloud providers through auditable Service Level Agreements (SLAs) [134]. To manage ongoing operational exposure, they institute XR-aware incident-response playbooks, run periodic red-teaming (including adversarial-ML tests), and monitor bot swarms and avatar takeovers in e-participation venues [137–139]. To deliver accountability without over-collection of personal data, they record credential/model-provenance in permissioned blockchain registries while keeping personally identifiable data off-chain.

Collectively, these countermeasures address GAN-enabled impersonation, third-party data-sovereignty constraints, and theft of credentials, identities, or avatars in urban-metaverse deployments, while directly mitigating the cost, compliance, edge-exposure, accessibility, and trust deficits surfaced earlier [137–139]. City deployments illustrate this layered approach: planning visualizations and city-scale DTs in London and Helsinki provide contexts where content-provenance controls are recommended against deepfakes [73,137,139]; Singapore operates under data-protection regimes with DPIA-like obligations for sensitive processing [134]; and applying zero-trust with device attestation is recommended for municipal XR endpoints, consistent with large-scale platforms such as Metaverse Seoul [73,136,139].

Targeted outreach, digital-literacy programs, and transparent accountability mechanisms are essential to build confidence in metaverse applications, ensuring they serve democratic ends rather than enable exclusion or manipulation.

Taken together, these challenges and risks as well as the corresponding countermeasures call for a coordinated approach that couples technical controls with ethical, legal, and organizational safeguards [3]. The themes above are synthesized in **Table 4**, which organizes challenges, risks and countermeasures across SC domains and highlights priority areas for policy and design intervention. Consistent with PRISMA guidance, **Table 4** was generated via a structured thematic analysis of the final set of included studies [140]; after identification, screening, and eligibility checks, recurring issues were coded and grouped by SC domain to condense complex evidence into a clear comparative format. In turn, the table deepens the analysis while making the challenges, risks and measures more accessible for policymaking and strategic planning.

**Table 4.** Mapping of key challenges, risks, and countermeasures to metaverse-SC implementation

| SC domains | Challenges | Risks | Countermeasures |
|---|---|---|---|
| **Smart Economy** | **Ch1**. Budgetary constraints and limited financial resources [129,137].<br>**Ch2.** High cost of XR infrastructure, devices, and platforms [14,129,130].<br>**Ch3.** Variability in public investment across cities, creating uneven urban development [3,131,135].<br>**Ch4**. Unclear legal status of digital assets, NFTs, and virtual transactions [14,134].<br>**Ch5**. Market access limitations for smaller enterprises due to high costs and platform dependencies [135].<br>**Ch6.** Skill gaps in the XR/AI workforce [3].<br>**Ch7.** Labor rights concern in the metaverse economy [130].<br>**Ch8.** High operational and energy costs of data centers supporting metaverse workloads [14,136]. | **R1.** Market monopolization suppresses competition and innovation [129, 135].<br>**R2**. Weak regulation of digital marketplaces that enables exploitative practices [3,134].<br>**R3.** Job displacement and labor insecurity [130].<br>**R4.** Financial damage from cyber incidents (e.g., ransom payments, recovery, and legal penalties) [137].<br>**R5**. Overreliance on commercial infrastructures that undermines public control [3,135]. | **Ch1-Ch3, Ch5, R1** → Phased rollouts and pilots [129].<br>**Ch1-Ch3, Ch5, R1** → Public-private cost-sharing [129].<br>**Ch1-Ch3, Ch5, R1** → Subsidized access and device lending [14,16].<br>**R2** → Open standards and interoperability clauses [129].<br>**Ch4** → Adoption of comprehensive legal frameworks tailored to SC contexts [134].<br>**Ch5, Ch6, R2** → City-wide upskilling/reskilling programs [14,16].<br>**Ch7, R4** → Cybersecurity frameworks, response protocols, and redress mechanisms [129,135,139].<br>**Ch8, R5** → Green-by-design infrastructure and renewable-powered data centers [14,136]. |
| **Smart People** | **Ch1.** Digital literacy gaps across demographics (e.g., elderly, migrants, low-income groups) [3,14,16,129,130,135].<br>**Ch2.** Language-restricted platforms reduce inclusivity (e.g., CatVers) [73].<br>**Ch3.** Limited public awareness of cybersecurity threats and metaverse risks. [73,129,137,138].<br>**Ch4.** Limited public understanding of benefits, implications, and responsible use [130,133].<br>**Ch5.** Lack of trust due to prior misuse, surveillance, or breaches [132]. | **R1.** Embedded discrimination from biased data or opaque processes [3,14,16,129,130,133,135,139].<br>**R2.** Psychological strain and digital escapism due to overexposure [135].<br>**R3.** Erosion of autonomy and dignity due to surveillance [130,132,133].<br>**R4.** Avatar hijacking and identity theft [73,129,138].<br>**R5.** Deepfakes distort discourse and learning integrity [137].<br>**R6.** Misuse or misunderstanding due to fear or lack of education [129,133]. | **Ch1, Ch3, Ch4, R1, R6** → Digital literacy campaigns, civic awareness, and public education about metaverse risks and rights [3,14,101,129,132,133].<br>**Ch2, R1** → Accessibility by design [135].<br>**R2** → Age-appropriate gates [129].<br>**Ch5, R3** → Policies for ethical development and surveillance protection [14,101,133].<br>**Ch6, R6** → Citizen co-creation and participatory governance [133]. |

Continuation Table:

| SC domains | Challenges | Risks | Countermeasures |
|---|---|---|---|
| | **Ch6.** Slow societal adoption and resistance to metaverse platforms [129].<br>**Ch7**. Mental-health challenges due to over-immersion and addictive behaviors [135]. | **R7.** Emotional fatigue and social isolation caused by excessive exposure [135].<br>**R8.** Sophisticated deepfakes (via GANs) threaten authenticity and trust [137]. | **R4, R5, R8** → DIDs/VCs, passkeys, provenance controls, and deepfake-detection pipelines [73,129,130].<br>**R5** → Hybrid deep-learning systems for GAN-based deepfake detection in education [137].<br>**Ch7, R7.** Mindfulness and digital-wellbeing programs promoting balance metaverse use [135]. |
| **Smart Governance** | **Ch1.** Exclusion of underrepresented groups in decision-making [3,129,131,135].<br>**Ch2.** Uneven public investment leads to spatial-social disparities [44,130,132].<br>**Ch3.** Dominance of private actors/ platforms in metaverse design [3,130,135].<br>**Ch4**. Non-inclusive interfaces excluding marginalized users [73,129].<br>**Ch5**. Fragmented or absent legislation on metaverse privacy and cybercrimes [129,134,136].<br>**Ch6.** Legal uncertainty regarding digital assets and platform interoperability [73,134].<br>**Ch7.** Lack of transparency and accountability in digital public services [129,130,132,134].<br>**Ch8**. Poor collaboration among stakeholders (e.g., government, private sectors, technology providers, and civic institutions) [133,137,139].<br>**Ch9.** Undefined legal accountability for failures or accidents (e.g., AVs) [136].<br>**Ch10.** Absence of robust cybersecurity strategies [137].<br>**Ch11.** Limited institutional capacity and shortage of skilled governance personnel for XR governance [129,133]. | **R1.** Bot/Sybil attacks distort e-participation and decision legitimacy [132,137,139].<br>**R2.** GDPR non-compliance and data protection violations [136,139].<br>**R3.** Deepfakes and impersonation erode trust in governance integrity [136–138].<br>**R4.** Weak identity verification enables manipulation [137].<br>**R5.** Surveillance misuse for control and profiling [3,14,132,133,134].<br>**R6.** Algorithmic opacity undermines procedural fairness [134].<br>**R7.** Disorganized data collection without informed consent [129,130,132,136,139].<br>**R8.** Public disengagement from civic participation [129].<br>**R9.** Centralized control over civic systems [133].<br>**R10.** Loss of public trust due to surveillance and lack of transparency [133,137].<br>**R11.** Cyberattacks on critical urban infrastructure (e.g., transportation, energy, or health services) [134,136].<br>**R12.** Autonomous weapons and AI warfare risks [133].<br>**R13.** Inconsistent enforcement of data-protection and cybersecurity laws across jurisdictions [134]. | **Ch1, Ch4, Ch8, R8** → Community participation, human-centered planning, and co-creation [132,133]<br>**Ch7, R7, R10** → Transparency portals, public reporting, and informed consent models [3,101,132,133].<br>**Ch5, Ch6, R2, R6, R7** → DPIAs, privacy-by-design, data minimization, and audit trails [129,132,137].<br>**Ch2, Ch3, R9** → Localization rules, vendor SLAs, and sovereignty frameworks [129,134,139].<br>**R1, R3, R4** → Bot/Sybil detection, identity verification, and credential provenance [132,134,136,137,139].<br>**Ch9, R12** → Legislative updates and accountability frameworks for AVs and AI warfare [133,134].<br>**Ch10, R11** → Zero-trust protocols, remote attestation, and XR-aware incident playbooks [16,136].<br>**Ch8, R9** → Multistakeholder collaboration [132,133,137].<br>**R2-R6** → Compliance with GDPR and other data protection standards [132,134,136].<br>**Ch11, R13** → Inter-agency coordination bodies for immersive-governance oversight and cross-border regulatory alignment [133,134]. |

Continuation Table:

| SC domains | Challenges | Risks | Countermeasures |
|---|---|---|---|
| **Smart Mobility** | **Ch1.** High infrastructure and network demand for transport twins [14,16,73,107].<br>**Ch2.** Need for real-time edge computing and data integration [129].<br>**Ch3.** Legal ambiguity around mixed autonomy liability [16,136].<br>**Ch4.** Exclusion of smaller cities due to unequal digital capacity [46]. | **R1**. SLAM data leakage exposes sensitive urban layouts [129,134,137].<br>**R2.** Sensor spoofing and telemetry hijacking [137,138].<br>**R3.** Cloud dependency increases data sovereignty risks [129,132].<br>**R4.** Continuous location tracking violates privacy [3,14,139].<br>**R5.** Liability disputes in AV accidents with mixed autonomy [136]. | **R1, R2** → SLAM minimization (on-device redaction) and pseudonymous telemetry [137].<br>**Ch1, Ch2, R2, R4** → Zero-trust networking, encryption, and session hardening [137].<br>**Ch3, R3** → Safety-case documentation and incident playbooks [16,136].<br>**Ch1-Ch3** → Accessibility by design in transport infrastructure [14,16].<br>**Ch4, R5** → Policy frameworks for equitable digital infrastructure and legal harmonization for AV responsibility [16,136]. |
| **Smart Environment** | **Ch1.** Metaverse deployments clash with privacy norms or protected areas [129].<br>**Ch2.** High energy demand of data centers, simulations, and edge workloads [14,136].<br>**Ch3.** Upfront sensing, mapping, and navigation investments [14].<br>**Ch4.** E-waste from rapid device cycles [129].<br>**Ch5.** Unequal distribution of environmental benefits [129].<br>**Ch6.** Misuse in smart energy systems [129].<br>**Ch7.** Lack of environment-related SC legislation [129].<br>**Ch8.** Lack of environmental impact assessments for immersive-technology deployments [129]. | **R1.** Spatial scanning compromises bystanders and environmental privacy [129,134].<br>**R2.** SLAM data leakage exposes infrastructure [129,134].<br>**R3.** Carbon footprint increases if optimization fails [3,14].<br>**R4.** Environmental injustice−concentration of harms in vulnerable areas [129].<br>**R5.** Regulatory non-compliance with environmental targets [129,136].<br>**R6.** Natural disaster vulnerabilities [129].<br>**R7.** Overdependence on unregulated sensing and data pipelines [129,134]. | **Ch2, Ch6, R3** → Carbon and energy reporting; renewable-by-default workloads [14, 129,136].<br>**Ch3, Ch4** → Device lifecycle and circularity policies [129].<br>**Ch5, R4** → Equity-by-design siting, benefits mapping, and community reviews [129].<br>**Ch1, R1, R2** → SLAM minimization and bystander protection through data redaction [129,134,137].<br>**Ch7, R5** → Environmental compliance frameworks and risk audits [129,136].<br>**R6** → Urban resilience plans addressing natural disasters [129].<br>**Ch8, R7** → Environmental risk audits and regulatory oversight for XR deployments [129,134]. |
| **Smart Living** | **Ch1.** Privacy concerns over data collection, use, and storage [131].<br>**Ch2.** Loss of social bonds due to algorithmic mediation [135].<br>**Ch3.** Cultural and educational access barriers due to cost [29,90,96].<br>**Ch4.** Poor understanding of safe use (safe-use practices) [132]. | **R1.** Digital escapism and compulsive behaviors harm well-being [135].<br>**R2.** Addiction lowers productivity and damages mental and social health [29,90,96].<br>**R3.** Opaque platforms obscure data use [132,133]. | **Ch1, R3, R5, R6** → Minimal data capture, secure storage and access, informed consent, safety protocols [132].<br>**Ch2, R1** → Safety-by-design, moderation, and reporting mechanisms [130,137,138].<br>**Ch3, R2** → Tiered pricing and access subsidies [29,90,96]. |

Continuation Table:

| SC domains | Challenges | Risks | Countermeasures |
|---|---|---|---|
| | **Ch5.** Exclusion of older or less tech-savvy populations from virtual healthcare [121,125].<br>**Ch6.** Over-commercialization of cultural-heritage content reducing authenticity [68,111]. | **R4.** Exposure to harassment, crimes, and data breaches [130, 132,137,138].<br>**R5.** Telemedicine data exposure or misuse [132].<br>**R6.** Limited awareness of consent and safety protocols [132].<br>**R7.** Inequitable access to virtual health and education services [121,125].<br>**R8.** Cultural dilution and loss of authenticity through virtual commercialization [68,111]. | **Ch4, R4, R5** → Content provenance systems and deepfake detection [138,139].<br>**Ch5, R7** → Hybrid physical-virtual healthcare and education models; accessibility by design in service delivery [121,125].<br>**Ch6, R8** → Cultural-integrity safeguards and balanced on-site/ virtual tourism policies [68,111]. |
| **Computing and Immersive Technology** | **Ch1.** Need for high-performance computing and technical capacity [3,71,129,130,135,136].<br>**Ch2.** Poor last-mile connectivity [135].<br>**Ch3.** Larger attack surface due to multi-device usage [129, 132,138].<br>**Ch4.** Cloud/edge reliance increases systemic complexity [129,132].<br>**Ch5.** Engineering burden for privacy-preserving ML (e.g., DPPML, secure-aggregation) [125,126].<br>**Ch6.** Lack of backup and recovery strategies for technical failures [132].<br>**Ch7.** High carbon footprint of large-scale AI/DT workloads [14,136]. | **R1.** Adversarial ML/sensor spoofing compromise XR perception [137].<br>**R2.** GAN-enabled deepfakes lead to impersonation and social engineering [136–138].<br>**R3.** Bot swarms and session hijacking [137].<br>**R4**. Telemetry replay corrupts immersive-session integrity [137,138].<br>**R5.** Cross-border processing violates data sovereignty [137,139].<br>**R6.** Cybersecurity protocol gaps expose urban systems [129,130,132,134].<br>**R7.** Behavioral/biometric data leakage (gaze, gait, motion) [132,128,139].<br>**R8.** Sensor noise misclassifies users or lets adversaries bypass systems [138].<br>**R9.** Lack of secure storage and sharing protocols [129,130].<br>**R10.** Sustainability failure if optimization and energy efficiency are not enforced [3,14]. | **Ch2, Ch4, R1-R6**. DIDs/ VCs passkeys, and provenance controls [130,137,139].<br>**Ch1, Ch2, R5, R6** → Zero-trust architecture, trusted XR endpoint management, and encryption [132].<br>**Ch3, R1, R4, R7** → Telemetry protection, adversarial ML defense, and behavioral privacy safeguards [132].<br>**Ch5, R1-R7** → DPPML frameworks: federated learning, secure aggregation, and differential privacy [125,126,132, 136,139].<br>**Ch4, R2, R3** → Bot/avatar-takeover monitoring, provenance recording, and remote attestation [130,134,137–139].<br>**Ch6, R9** → Backup strategies, secure data storage, and sharing protocols [132,137].<br>**Ch7, R10** → Carbon and energy monitoring, renewable-by-default computing, and lifecycle optimization [14,136]. |

Addressing these challenges and mitigating the associated risks requires a coherent framework that anchors metaverse integration in trust, accessibility, and governance. CitiVerse represents this framework, translating strategic safeguards into actionable, city-centric enablers. The transition from the broader metaverse-SC framework to the CitiVerse enablers reflects the need to refine generalized digital transformation trends into practical, city-centric applications that directly enhance urban life. While the metaverse drives macro-level economic, technological, and industrial innovation, CitiVerse concentrates on localized, citizen-driven experiences, ensuring active participation, service accessibility, and municipal collaboration. This selection process is based on thematic extraction and urban relevance, prioritizing elements that empower citizens, strengthen governance, and facilitate inclusive digital services.

Global initiatives and ITU reports emphasize citizen co-creation, AI-driven service integration, and accessibility as critical factors for urban digitalization. Accordingly, the CitiVerse framework bridges the gap between high-level metaverse capabilities and tangible, everyday SC interactions, ensuring that technology serves urban communities in a participatory, inclusive, and governance-focused manner.

Key CitiVerse enablers:

- **Active citizen participation:** CitiVerse enables residents to contribute ideas and feedback on infrastructure projects and community initiatives, fostering a sense of ownership and accountability.
- **Use of avatars:** Citizens use avatars to navigate and interact across digital and physical spaces, enhancing engagement with urban environments and supporting personalized participation [9].
- **Innovative service delivery:** CitiVerse opens avenues for businesses to offer immersive, resident-oriented services, such as virtual storefronts and experiential engagement [56].
- **Collaboration with local governments:** Businesses can co-develop services with municipalities (e.g., virtual health consultations, educational programs, public service announcements) aligned with local needs [6].
- **Breaking down barriers:** CitiVerse reduces access barriers through inclusive virtual environments that support universal participation by people with disabilities and marginalized communities [3].
- **Remote access to services:** CitiVerse facilitates remote access to education, healthcare, and government resources, simplifying resident interaction with local authorities [95].

The ITU [57,58,141], acknowledging these advancements, has established dedicated Focus Groups to address technical, ethical, and policy challenges associated with CitiVerse implementation under the initiative “Shaping a CitiVerse for All.” Their recommendations emphasize the need for interoperability between digital and physical infrastructures (to prevent fragmentation) [10,11,12], robust cybersecurity and data privacy frameworks (to mitigate sensor-rich and biometric risks while enabling transparent governance) [132–139], and universal accessibility through inclusive design (to tackle device, skills, and disability barriers) [130,138,139]. They further emphasize public-private collaboration with clear accountability to enhance governance capacity and sustainability and citizen-engagement strategies to build trust, legitimacy, and meaningful participation [131]. In short, ITU guidance positions CitiVerse as the operational layer through which cities implement essential governance and technical controls while keeping deployments people-centric.

This metaverse-enabled transformation equips SCs with practical levers for implementation. Specifically, DTs allow policies to be simulated and evaluated before deployment, remote-by-design access broadens the reach of education, healthcare, and administrative services, while structured multi-stakeholder collaboration supports participatory planning and decision-making. In parallel, it unlocks new forms of economic activity, from immersive services to creator-driven digital assets, while ensuring that implementations remain anchored in privacy, security, accessibility, and accountability.

Overall, CitiVerse addresses risks such as privacy, deepfakes, algorithmic bias, and digital exclusion by embedding provenance verification, zero trust security, DPPML, accessibility by design, and accountable governance [125,126]. These enablers empower cities to pilot, evaluate, and scale trustworthy metaverse-based services across urban domains.

**Figure 3** illustrates the 6 pillars of SCs alongside CitiVerse's core dimensions, showing how these enablers interoperate across domains to support holistic, inclusive, and sustainable urban ecosystems.

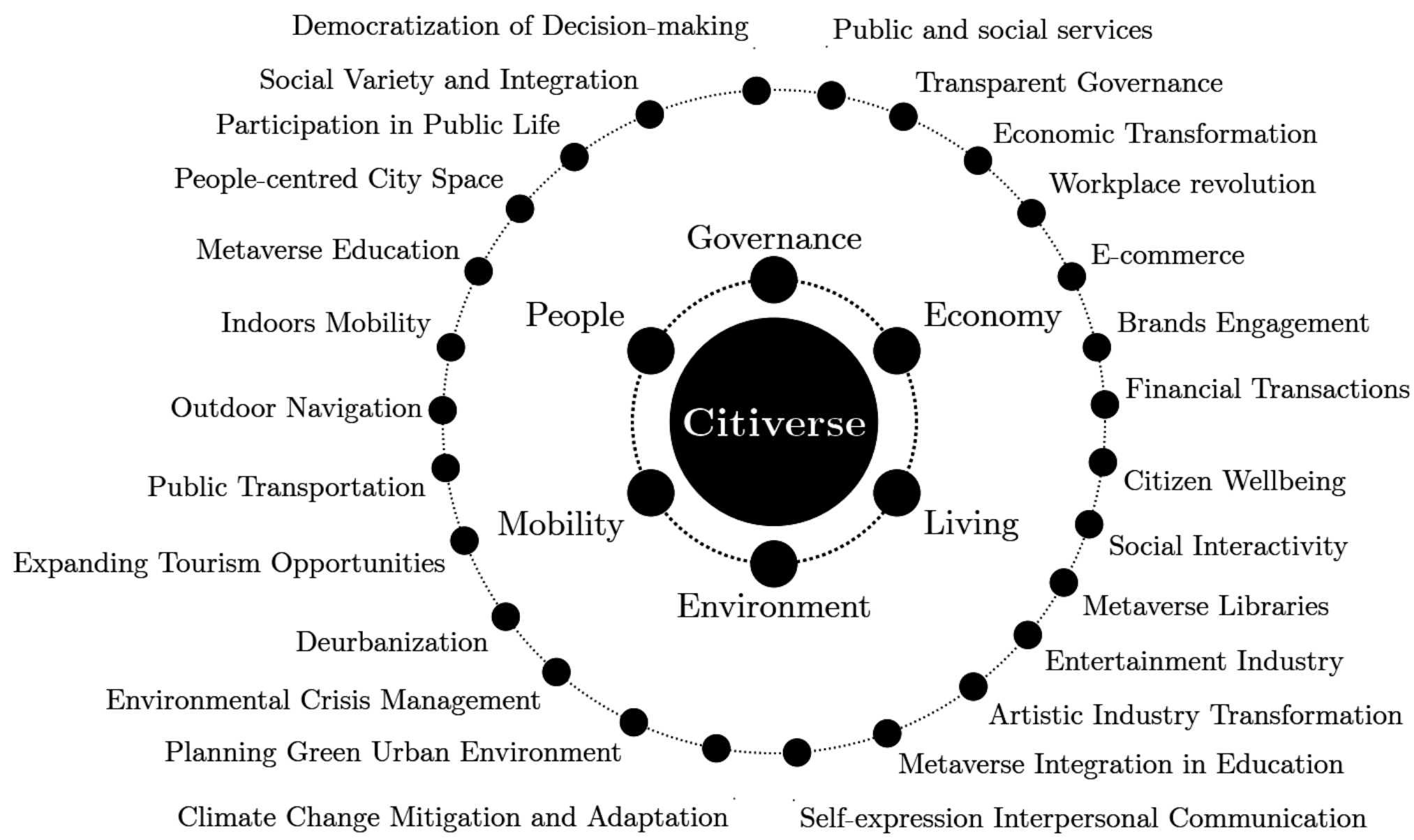

**Figure 3.** The six pillars of the SC in accordance with the core dimensions of the CitiVerse

## 5.2. Critical synthesis: Metaverse-SCs SWOT analysis

The integration of metaverse technologies into SCs presents a dynamic interplay of opportunities and risks. While enhancing urban innovation, citizen engagement, and economic diversification, the metaverse also introduces challenges related to privacy, digital literacy, and infrastructure costs. A structured SWOT analysis [142] (**Table 5**) provides an in-depth evaluation of the key factors influencing metaverse adoption in SCs, offering insights into how cities can leverage strengths, mitigate weaknesses, capitalize on opportunities, and address potential threats in the pursuit of sustainable, inclusive, and trustworthy digital urban ecosystems.

**Table 5.** The SWOT analysis for SCs and the metaverse

| **Strengths** | |
|---|---|
| Economic diversification: The metaverse enables new business models via virtual goods, immersive commerce, and innovative financial tools such as cryptocurrencies and NFTs. | [21,54,77,79,80,84,86,143] |
| Enhanced productivity: Virtual workplaces and globally distributed teams maximize global talent utilization and reduce operational barriers. | [3,12,77] |
| Educational innovation: Immersive learning and Vocational Education and Training (VET) platforms enhance skill development and lifelong learning. | [13–15,62,85,94,96] |

Continuation Table:

| | |
|---|---|
| Increased accessibility: Real-time translation and inclusive virtual classrooms reduce geographical and language barriers. | [13–15,56,59,97,98] |
| Participatory decision-making: Metaverse platforms enable real-time citizen engagement, citizen co-design, and policy co-creation. | [3,44–46,100,102–104] |
| Transparency: Metaverse-supported governance can improve accountability through open information and auditable visualizations. | [47,103,134] |
| Green mobility: Metaverse tools can optimize traffic management, public transport, and AV deployment. | [9,78,108,110–113] |
| Enhanced safety: Context-aware navigation and simulation reduce accidents and improve emergency response. | [107,114] |
| Sustainability: Simulation-driven planning and resource optimization support greener and more resilient cities. | [3,115] |
| Crisis management: Predictive tools increase preparedness for environmental disasters. | [67,114–117] |
| Improved well-being: Applications in healthcare, entertainment, and social interaction support mental and social health. | [68,74,88,109,118,121–123] |
| Inclusivity: Metaverse technologies expand participation in social life and access to information for marginalized groups and people with disabilities. | [3,6,46,88] |
| City micro-cases:<br>Seoul's virtual services and Dubai's immersive branding illustrate economic diversification. Helsinki shows productivity gains through remote collaboration.<br>London demonstrates more transparent planning via AR/DT visualizations.<br>Singapore and London frequently pair DT layers with operations to improve mobility, safety, sustainability, crisis readiness, well-being, and inclusion. | [73,129,130] |
| **Weaknesses** | |
| Technological and economic barriers: High infrastructure and development costs constrain adoption in less-resourced regions. | [14,15,17,18,129,130–135] |
| Skills gap: Limited digital literacy and XR expertise hinder uptake and talent attraction. | [3,14,16,46,119,130–135] |
| Fragmented implementation: Poor interoperability between metaverse platforms and SC systems reduces efficiency. | [47,135] |
| City micro-cases:<br>Device affordability constraints in Seoul and platform licensing in London create budget pressure. Skills gaps in Helsinki slow uptake despite upskilling/reskilling; and noninteroperable stacks reduce efficiency. | [73,129] |
| **Opportunities** | |
| Global collaboration: The metaverse enables cross-border partnerships in governance, education, and commerce. | [6,13–15,19,46,68,77,] |
| Personalized services: Tailored virtual environments support personalized services in critical areas such as health, education, and safety. | [56,68,74,85,88,109,118,121–123] |
| Data-driven decision-making: Combining virtual-world telemetry with SC data improves planning across domains. | [93] |
| Urban decongestion: Remote participation and telepresence can reduce trips and peak-hour load. | [3,116] |
| Inclusive development: Co-design and participatory tooling foster stronger communities. | [45] |

Continuation Table:

| | |
|---|---|
| City micro-cases:<br>Singapore-Nordic exchanges on DT standards illustrate collaboration.<br>Seoul's virtual counters show personalized services.<br>London's 3D models and city data strengthen planning; and telepresence for permits/hearings broadens participation and reduces travel. | [73,129,134] |
| **Threats** | |
| Privacy and security risks: Data collection and breaches undermine trust in metaverse environments. | [3,90,93,132–139,144] |
| Ethical challenges: Surveillance concerns, data misuse, and attention/addiction risks can harm social cohesion. | [3,5,6,7,65,133] |
| Regulatory uncertainty: Ambiguity on ownership, digital assets, governance, and intellectual property complicates deployment. | [14] |
| Economic instability: Volatility in digital assets (e.g., cryptocurrencies) and overreliance on virtual economies increase exposure. | [54,] |
| Resistance to adoption: Public skepticism, especially among older or non-tech-savvy users, can slow progress. | [14,101] |
| City micro-cases:<br>Public-space analytics in London and centralized data stacks in Singapore raise trust concerns.<br>Rapid rollouts in Dubai can outpace ethics oversight; unsettled digital-asset/data rules increase regulatory risk; low digital trust in older/non-tech-savvy cohorts threatens uptake trajectories. | [73,134,130] |

### 5.3. Key contributions of the study: Positioning and novelty

This study makes a clear, decision-oriented leap in the metaverse-SC conversation by turning a diffuse technological promise into a concrete, city-ready playbook that leaders and researchers can actually use.

• **Conceptual and theoretical contribution:** The study introduces and develops CitiVerse, a citizen-centric reference model that maps core metaverse capabilities, such as simulation, immersion, co-creation, and data-driven services, onto Giffinger's 6 pillars (Smart Economy, Smart People, Smart Governance, Smart Mobility, Smart Environment, Smart Living). In doing so, it shifts from a technology wishlist to a service-orientated urban blueprint, clarifying how immersive, data-driven functions align with established SC dimensions.

• **Methodological contribution:** It combines a PRISMA-based systematic review with a comparative cross-case analysis, including Seoul, Dubai, Shanghai, Helsinki, London, New York City, Boston, Tampere, Catalonia, Barbados. This reproducible design moves beyond narrative surveys to explain why outcomes diverge across governance models and data-rights regimes, yielding a rigorous, multi-angle synthesis rarely achieved in prior research.

• **Empirical and practical contribution**: It consolidates how cities pilot metaverse applications across economic development, participation and service delivery, mobility and navigation, culture and tourism, and environmental management. From this evidence, it distills actionable guidance for municipal leaders and urban designers, which could operate as a strategic roadmap for responsible, inclusive, and effective public-sector adoption.

- **Novelty relative to prior literature:** Whereas most prior work is education- or consumer-centric, this study directly addresses the urban-integration problem, linking metaverse technologies to SC governance and infrastructure, and positions CitiVerse as a reusable reference model for researchers and practitioners who seek to evaluate or design metaverse-enabled public services.

Overall, the study treats the SC and the metaverse as one integrated system rather than parallel domains. Its contribution is a decision-oriented synthesis: it pinpoints where the metaverse adds public value (participatory planning, accessible services, simulation-driven optimization) and where safeguards are essential (privacy, data sovereignty, inclusion).

For city leaders and planners, the roadmap is clear: Align pilots with the 6 pillars, embed accessibility-by-design and robust security controls, and track outcomes with transparent, citizen-facing metrics. For researchers, it sets a forward agenda: Comparative evaluations of governance models, longitudinal assessment of social and environmental impacts, rigorous participation/trust metrics, and advances in privacy-preserving analytics and interoperability for city contexts.

In sum, the study blends conceptual clarity, methodological rigor, and practice-ready guidance, charting a path toward metaverse–SCs that are resilient, inclusive, and genuinely citizen-centric.

## 6. Conclusion

This study examines how metaverse capabilities can practically reshape SCs by introducing CitiVerse as a city-centric and citizen-first operational layer. It synthesizes the literature through a PRISMA-based systematic review and a comparative analysis of leading city cases to show where immersion, simulation, and data-driven services add public value across Giffinger's 6 pillars. It clarifies core concepts and functional roles, including XR for interaction, DTs for evidence-based simulation, provenance and zero-trust models for security, and inclusive design for accessibility, and connects them to concrete governance mechanisms such as DPIAs, data minimization, and transparent oversight. It also maps the landscape of challenges and risks related to cost, skills, privacy, interoperability, and trust, within a layered control architecture encompassing identity, content integrity, network and edge security, analytics, and governance. In doing so, it transforms a diffuse debate into an operational framework that cities can pilot, evaluate, and scale responsibly.

This study differs from prior research by moving beyond conceptual discussion toward practical implementation. While most studies focus on education, gaming, or consumer experiences, it examines how metaverse technologies can be effectively integrated into city governance and public services. By applying the CitiVerse model as a guiding framework, the study provides policymakers, planners, and researchers with a concrete reference for understanding where immersive tools create public value and how their associated risks can be responsibly managed. By linking technological, ethical, and social dimensions, it helps cities design and test pilots that are inclusive, secure, and sustainable. Its findings support evidence-based decision-making, strengthen public trust, and guide future research toward measurable and citizen-centered outcomes.

The study's main limitation lies in its reliance on secondary data from published literature and city reports, which may not fully capture rapidly evolving metaverse practices or proprietary municipal projects. The comparative case analysis focuses on a limited number of leading cities, which may constrain generalizability to smaller or less digitally mature contexts. In addition, technological and regulatory developments are advancing more rapidly than academic evaluation, meaning that some findings may become outdated as standards and infrastructures evolve. Self-reported municipal data may also introduce optimism or selection bias, and language or regional coverage could skew results toward better-documented cities. Finally, limited public data on costs, contracts, and security incidents restricts detailed verification of claims.

Future research should build on this study by testing the CitiVerse framework in real urban contexts and measuring its social, technical, and governance outcomes. Pilot projects in collaboration with municipalities could explore how immersive participation, DTs, and data-protection tools perform in everyday city operations. Further studies should develop practical metrics for inclusion, transparency, and trust, ensuring that digital transformation remains citizen-centered. Researchers could also investigate the long-term effects on sustainability, equality, and public value, while expanding interdisciplinary cooperation among urban planning, data science, and public administration to translate CitiVerse from concept to practical urban innovation.

**Author contributions:** All authors contributed equally to the manuscript and read and approved the final version of the manuscript. More specifically, Conceptualization, P. Fitsilis, I. Chatzopoulou and P. Tsoutsa; Methodology, I. Chatzopoulou and P. Tsoutsa; Validation, P. Fitsilis and P. Tsoutsa; writing—original draft preparation, I. Chatzopoulou; writing—review and editing, P. Toutsa; supervision, P. Fitsilis.

**Conflict of interest:** The authors declare no conflict of interest.